\documentclass[a4paper,11pt]{article}

\pdfoutput=1 

\usepackage{jheppub_modified} 

\usepackage[T1]{fontenc} 

\usepackage{verbatim}
\usepackage{color}
\usepackage{graphicx}
\usepackage{subcaption}
\usepackage{bigcenter}

\usepackage{sectsty}
\allsectionsfont{\boldmath}

\usepackage{xspace}
\usepackage[dvipsnames,table]{xcolor}
\usepackage{tabularx}
\usepackage[normalem]{ulem}
\newcolumntype{C}[1]{>{\centering\arraybackslash}p{#1}}

\newcommand{\eq}[1]{Eq.~(\ref{#1})}
\newcommand{\bib}[1]{Ref.~\cite{#1}}

\newcommand{\fig}[1]{Fig.~\ref{#1}}
\newcommand{\tab}[1]{Table~\ref{#1}}

\newcommand{\sect}[1]{Section~\ref{#1}}

\newcommand{\appen}[1]{Appendix~\ref{#1}}

\newcommand{\bea}{\begin{eqnarray}}
\newcommand{\eea}{\end{eqnarray}}

\newcommand{\crn}{\nonumber \\}

\newcommand{\fr}{\frac}

\newcommand{\abs}[1]{\left\vert#1\right\vert}

\newcommand{\gev}{{\unskip\,\text{GeV}}}

\title{Triply polarized $WWW$ at the LHC: first glimpse at LO}

\author[]{Van Cuong Le,}
\author[]{Duc Ninh Le,}
\author[]{Thi Nhung Dao}

\affiliation[]{Phenikaa Institute for Advanced Study, Phenikaa University, Hanoi 12116, Vietnam}

\emailAdd{cuong.levan@phenikaa-uni.edu.vn}
\emailAdd{ninh.leduc@phenikaa-uni.edu.vn}
\emailAdd{nhung.daothi@phenikaa-uni.edu.vn}

\abstract{We present first results for triply polarized $WWW$ events at the LHC. The calculation 
is performed at leading order for fully leptonic decays using the Standard Model (SM). Employing an inclusive kinematic 
cut setup, we found that the triply-transverse polarization fraction is about $51\%$, while the triply-longitudinal (LLL) fraction is smallest with $1.4\%$ for 
the $W^-W^+W^+$ process. 
The interference between different polarization amplitudes amounts to $+1.8\%$. Results for the $W^+W^-W^-$ case are similar.
On the technical side, a new general on-shell mapping for multi-resonance processes, being a 
crucial element of polarized cross-section calculation, is also presented. 
Finally, comparisons with the results of MoCaNLO and Sherpa 
are provided, showing good agreements.}

\begin{document}
 \maketitle
\flushbottom

\section{Introduction}
\label{intro}
Finding new physics beyond the Standard Model (SM) is not easy. Precision is really the key in this effort. 

The Large Hadron Collider (LHC), where two bunches of protons are collided at $13.6$ TeV, is a great playground to understand the 
boundaries of particle physics. The ATLAS and CMS detectors have been able to distinguish the top quarks, massive gauge bosons, and 
Higgs boson, produced in the collisions. Moreover, different polarized states of the massive gauge bosons have been 
observed in diboson processes \cite{CMS:2021icx,ATLAS:2024qbd,ATLAS:2025wuw}. 
These measurements of polarized cross sections are based on 
template-fitting methods, where polarization templates are produced by simulation. 
Polarization and quantum entanglement observables of top-quark pairs have also recently been measured \cite{ATLAS:2023fsd,CMS:2024zkc}.

Differently from the case of diboson production, triboson processes offer new polarization observables, in particular the triply longitudinally (LLL) 
polarized cross sections. As will be pointed out in this paper, the LLL cross sections are suppressed in the SM, hence any deviation between the 
SM prediction and data would hint to new physics. The triply transverse polarization and mixed cases can also provide additional information about the 
structure of the quartic gauge couplings or new physics. 
Moreover, triboson processes can probe the quartic gauge couplings in different kinematical 
regions compared to vector boson scattering processes. 
It is therefore evident that triboson production together with the inclusive diboson processes and 
$VVjj$ production will provide more complete information on the triple and quartic gauge couplings, 
the $HVV$ couplings \cite{Henning:2018kys}, the electroweak symmetry breaking mechanism, and unitary cancellation 
at high energies. Surprisingly, triboson processes are found to be sensitive to the light-quark Yukawa couplings via 
a rapid growth of the partonic cross sections with energy due to the Higgs-exchange diagrams \cite{Falkowski:2020znk}.  

Even though the cross sections of massive triboson production processes are very small, after 16 years of the LHC operation, 
measurements of these cross sections have been achieved by ATLAS \cite{ATLAS:2024nab} and CMS \cite{CMS:2020hjs,CMS:2025hlu}. 
The precision is of course still limited at the moment. However, with continuing efforts by experimentalists and theorists, 
we will undoubtedly reduce the errors to get hints of new physics.   

Knowing polarization templates with high precision is therefore important. For inclusive diboson processes, results at the level 
of next-to-next-to-leading order (NNLO) QCD and next-to-leading order (NLO) electroweak (EW) have recently been achieved (see \cite{Denner:2021csi,Le:2022ppa,Pelliccioli:2025com} for key information). Polarized vector boson scattering has been studied in \cite{Ballestrero:2017bxn} for $W^+W^-$ case, \cite{Ballestrero:2019qoy,Denner:2025xdz} for $WZ$, 
\cite{Ballestrero:2019qoy,Denner:2026vmf} for $ZZ$, and \cite{Ballestrero:2020qgv,Denner:2024tlu} for the same-sign $W^+W^+$ case.

In these works, polarized cross sections are calculated by using a technique 
named Pole Approximation (PA) \cite{Stuart:1991cc,Stuart:1991xk,Aeppli:1993cb,Aeppli:1993rs,Denner:2000bj}. 
This method has been advanced up to the level of NLO EW \cite{Denner:2021csi,Le:2022ppa}. 
Alternatively, the spin-correlated narrow-width approximation (NWA) \cite{Richardson:2001df,Uhlemann:2008pm,Artoisenet:2012st} 
has been used in Monte-Carlo event generators including MG5\_aMC \cite{Stelzer:1994ta,Alwall:2014hca,Frederix:2018nkq} at leading order \cite{BuarqueFranzosi:2019boy}
and Sherpa \cite{Sherpa:2024mfk} at an approximate NLO QCD \cite{Hoppe:2023uux}. 
Beyond these two on-shell approximations (which guarantee gauge invariance), a method based on resonant-diagram selection has been investigated in \cite{Javurkova:2024bwa,Carrivale:2025mjy} 
for the $ZZ$ production process using MG5\_aMC. As off-shell momenta are used in the polarized amplitudes, gauge invariance is violated. The level of violation must then be controlled by 
using kinematic cuts focusing on the poles of the resonances.

Parton shower is important for the improvement of the fixed-order results 
as well as the generation of events. This has been investigated for polarized diboson in \cite{Pelliccioli:2023zpd} at NLO QCD level using 
Powheg-Box-Res framework \cite{Nason:2004rx,Frixione:2007vw,Alioli:2010xd,Jezo:2015aia}, 
and for polarized multi-boson at an approximate NLO QCD \cite{Hoppe:2023uux}. MG5\_aMC can also be used 
for tree-level \cite{Carrivale:2025mjy} or loop-induced processes \cite{Javurkova:2024bwa}. 

The effects of higher-order operators using the SM effective field theory framework on polarization observables of 
multi-boson processes have been investigated in \cite{ElFaham:2024uop,Celada:2024cxw,Haisch:2025jqr,ElFaham:2025fow,Pelliccioli:2026ltl}. 
Moreover, machine-learning techniques to estimate polarized cross sections have been developed in \cite{Grossi:2023fqq} at LO and 
very recently in \cite{Cruz-Martinez:2026uud} at NLO QCD.

In this work, we take the first step for the calculation of polarized cross sections of the inclusive triboson processes. 
A needed ingredient is the on-shell mapping for the Triple Pole Approximation (TPA). A general and robust method, applicable for $n$-resonance 
processes with arbitrary number of decay products, is identified in this study. 
With this new method, first results at leading order (LO) for triply-polarized cross sections of the $WWW$ production processes at the LHC are presented in this work. 
Extension to NLO QCD+EW needs additional care due to many subtleties similar to the 
diboson processes \cite{Denner:2021csi,Le:2022ppa,Dao:2023kwc,Denner:2024tlu}. 
We are currently working on this and the results will be presented when available. 

Our presentation is outlined as follows. The definition of polarized cross sections using the TPA is given in \sect{sec:cal_med}. 
The new on-shell mapping is presented in \sect{sec:map_os} and details of our code implementation in \sect{sec:code}. 
Numerical results for the $W^-W^+W^+$ process are provided in \sect{sec:res}, while those of the $W^+W^-W^-$ case are in \appen{appen_pmm}. 
Comparisons with MoCaNLO and Sherpa are presented in \sect{sect:compa}.
Conclusions are given in \sect{sec:con}.
\section{Details of calculation}
\label{sec:cal_med}
We consider the following process:
\begin{align}
p + p \to l_1 + l_2 + l_3
+ l_4 + l_5 + l_6 + X,
\label{eq:process_pp_6l}
\end{align}
where the final state can be $e^- \bar{\nu}_e \mu^+ \nu_\mu \tau^+ \nu_\tau$ or $ e^+ \nu_e  \mu^- \bar{\nu}_\mu  \tau^- \bar{\nu}_\tau$. 
For the case of identical leptons such as $\mu^-\bar{\nu}_\mu e^+\nu_e e^+\nu_e$, ignoring the small interferences due to identical particles, 
the cross sections are obtained from the above different-flavor final states by multiplying the final results by a factor $1/2$. 
The sum over all final states (i.e. $e^-\mu^+\tau^+$, $\mu^-e^+\tau^+$, $\tau^-e^+\mu^+$, $e^-\mu^+\mu^+$, $e^-\tau^+\tau^+$, 
$\mu^-e^+e^+$, $\mu^-\tau^+\tau^+$, $\tau^-e^+e^+$, $\tau^-\mu^+\mu^+$ for the $W^-W^+W^+$ case, the corresponding neutrinos have been omitted for 
simplicity) can thus be achieved by multiplying our results by a factor $6$.

\begin{figure}[ht!]
	\centering
	\includegraphics[width=0.8\textwidth]{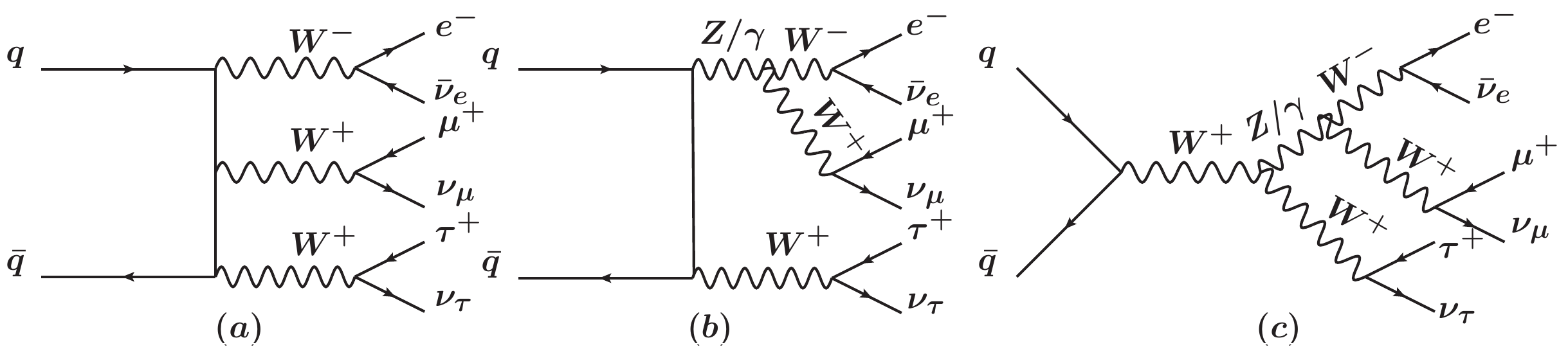}
	\includegraphics[width=0.8\textwidth]{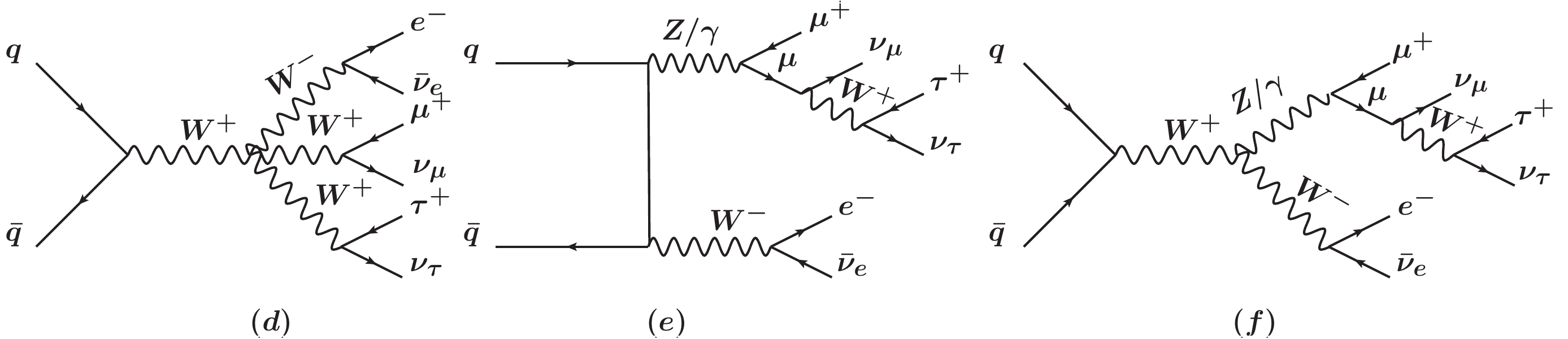}	
	\includegraphics[width=0.8\textwidth]{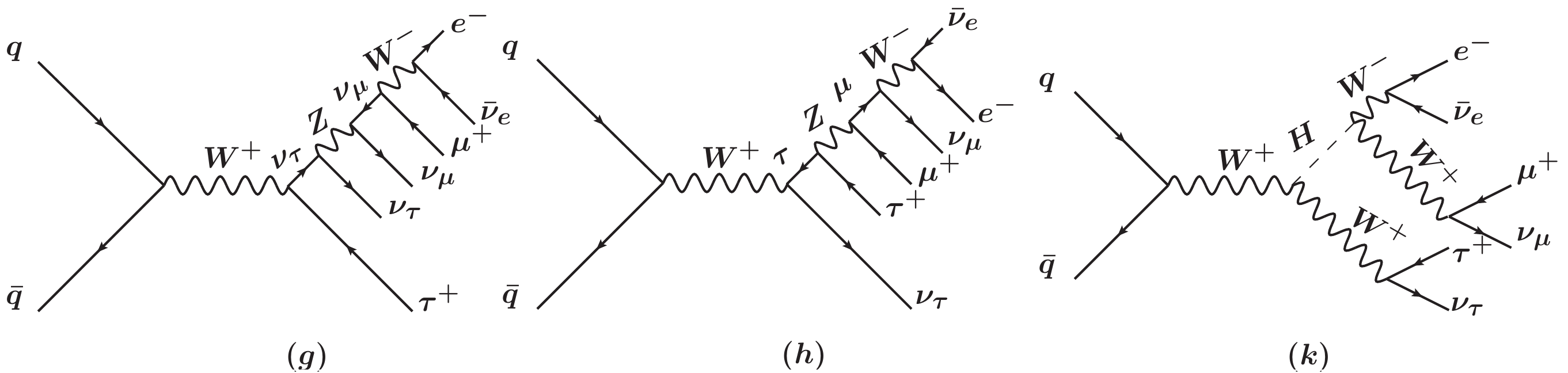}
	\caption{Representative Feynman diagrams at Born level for the full off-shell process, 
	classified into the triply-resonant diagrams (a, b, c, d, k), doubly-resonant diagrams (e, f), 
	singly-resonant diagrams (g, h), and $WH$ diagram (k).}
	\label{Feyn_dia}
\end{figure}

At LO, the contributing partonic processes are:
\begin{align}
\bar{q}(k_1) + q^\prime(k_2) \to l_1 (k_3) + l_2 (k_4) + l_3
(k_5) + l_4 (k_6) +  l_5 (k_7) + l_6 (k_8),
\label{eq:process_qq_6l}
\end{align}
with $q$ and $q^\prime$ being the quarks of the first two generations.
Representative Feynman diagrams at LO are shown in \fig{Feyn_dia}.

Among these 6-lepton events, we are interested in events coming from the polarized $WWW$ system. 
These events are produced by the following processes:
\begin{align}
\bar{q}(k_1) + q^\prime(k_2) \to V_1 (q_1) + V_2 (q_2) + V_3 (q_3)
\to  \sum_{i=1}^6 l_i (k_{i+2}),
\label{eq:proc_signal}
\end{align}
where the intermediate $V_1 V_2 V_3$ can be $W^-W^+W^+$ or $W^+W^-W^-$, corresponding to the final state
$e^- \bar{\nu}_e \mu^+ \nu_\mu \tau^+ \nu_\tau$ or $ e^+ \nu_e  \mu^- \bar{\nu}_\mu  \tau^- \bar{\nu}_\tau$, respectively. 
These signal processes will therefore be referred to as $W^-W^+W^+$ or $W^+W^-W^-$ for short. Representative Feynman diagrams 
of this mechanism are displayed in \fig{Feyn_dia} (a, b, c, d, k).

One observes that the diagrams belonging to this triply-resonant class, being a subset of the full process \eq{eq:process_qq_6l}, do not form a gauge-invariant group 
for a general kinematic configuration. However, gauge-invariant results can be obtained by summing over on-shell kinematic configurations. 
These configurations can be calculated by means of an on-shell mapping as specified in \sect{sec:map_os}. 
For these on-shell momentum configurations, denoted with a hat, we have: 
\bea
\hat{q}_i = M^2_{V_i},\quad i=1,2,3.
\eea
With the intermediate gauge bosons satisfying these on-shell conditions, we can then factorize the helicity amplitude for the process 
\eq{eq:proc_signal} into gauge-invariant factors, being the on-shell production amplitude $\bar{q} q^\prime \to V_1 V_2 V_3$ and 
three on-shell decay amplitudes $V_i \to l l^\prime$. Specifically, for a given helicity configuration of the initial-state quarks and final-state leptons, at leading order we have
\begin{align}
\mathcal{A}^{\bar{q}q^\prime \rightarrow V_1 V_2 V_3 \rightarrow 6l}_{\text{TPA}}([k_i]) 
= \frac{1}{Q_1Q_2Q_3}
\sum_{\lambda_1,\lambda_2,\lambda_3=1}^3 
&\Big[ \mathcal{A}^{\bar{q}q^\prime \rightarrow V_1 V_2 V_3} ([\hat{k}_i],\lambda_1,\lambda_2,\lambda_3)\mathcal{A}^{V_1 \rightarrow l_1 l_2} ([\hat{k}_i],\lambda_1)\crn
&\mathcal{A}^{V_2 \rightarrow l_3 l_4} ([\hat{k}_i],\lambda_2) 
\mathcal{A}^{V_3 \rightarrow l_5 l_6} ([\hat{k}_i],\lambda_3)\Big],
\label{eq:def_TPA}
\end{align}
where
\begin{align}
Q_i = q_i^2 - M_{V_i}^2 + i M_{V_i} \Gamma_{V_i} ~(i=1,2,3),
\end{align}
with the off-shell momenta $q_1 = k_3 + k_4$, $q_2 = k_5 + k_6$, $q_3 = k_7 + k_8$ 
and $M_{V_i}$, $\Gamma_{V_i}$ being the pole mass and the total decay width of the resonance $V_i$, respectively. 
The index $\lambda_i$ is the polarization (or helicity) index of the massive gauge boson $V_i$. 
$[k_i]$ is a set of off-shell external momenta, while $[\hat{k}_i]$ is a set of on-shell external momenta.
In practice, $[\hat{k}_i]$ is obtained from $[k_i]$ by using an on-shell mapping, see \sect{sec:map_os}. 
This mapping is not unique, but the results of different mappings have been found to be in good agreement at the 
integrated cross section and differential cross section levels. More details are provided in 
\sect{sec:map_os} and \sect{sect:compa}.
The TPA is a straightforward extension of the Double Pole Approximation (DPA) used 
for diboson production processes \cite{Aeppli:1993cb,Aeppli:1993rs,Denner:2000bj,Ballestrero:2017bxn}. 
Recently, the DPA has been extensively used to define polarized cross sections of diboson processes at NLO EW and NNLO QCD levels \cite{Denner:2021csi,Poncelet:2021jmj,Le:2022ppa}. 
The TPA has been employed in \cite{Dittmaier:2019twg} for the case of unpolarized $WWW$ production at the LHC.

From the TPA master equation \eq{eq:def_TPA} we then define the polarized cross sections as follows. 
For example, the LLL cross section is calculated by selecting the term with $\lambda_1 = \lambda_2 = \lambda_3 = 2$ 
in the r.h.s. of \eq{eq:def_TPA}. 
The TTT (triply transverse) one includes eight terms with $\lambda_i = 1$ or $3$ ($1$ for left transverse, $3$ for right transverse). 
Interference between these left-transverse and right-transverse modes are therefore included in the TTT cross section. 
Other polarizations including TTL and TLL (the position of the L or T is unimportant) are similarly defined as: 
\begin{align}
\mathcal{A}_\text{TTL} &= \sum_{\lambda_1,\lambda_2,\lambda_3\in [1,2,3|\exists! \lambda_i=2]} \mathcal{A}_\text{TPA}(\lambda_1,\lambda_2,\lambda_3),\\
\mathcal{A}_\text{TLL} &= \sum_{\lambda_1,\lambda_2,\lambda_3\in [1,2,3|\exists! \lambda_i\in [1,3]]} \mathcal{A}_\text{TPA}(\lambda_1,\lambda_2,\lambda_3).
\end{align}
Taking the square of the modulus of these polarized amplitudes includes interference between various individual polarization 
combinations, such as between $W^-_T$ $W^+_T$ $W^+_L$ and $W^-_L$ $W^+_T$ $W^+_T$ etc. While the unpolarized cross section is Lorentz invariant, polarized 
ones are not as the summation over the polarization indices is not complete. We therefore choose the $VVV$ center-of-mass frame (c.m.f.) to 
define polarizations in this study. Practically, this means that the on-shell momenta must be calculated in the $VVV$ c.m.f.. 
The off-shell denominators $Q_i$ in \eq{eq:def_TPA} are Lorentz invariant, hence 
can be calculated in any reference frame.

The TPA unpolarized cross section then reads
\begin{align}
\sigma^{\text{unpol}}_\text{TPA} 
= \sigma_{\text{TTT}} + \sigma_{\text{TTL}} + \sigma_{\text{TLL}} + \sigma_{\text{LLL}}
 + \sigma_{\text{Inf}},  \label{def_xs_unpol}
\end{align}
where the last term is the remaining polarization interference, which will be referred to as the polarization interference.   

There are two ways to calculate the interference, via direct integration of the amplitude-amplitude interference 
or as the difference of the unpolarized cross section and the sum of the polarized cross sections. 
The former is more precise but the Monte-Carlo integration is challenging due to cancellation between positive 
and negative contributions across different phase-space regions. For the scale-dependence plots 
in \fig{fig:xi_dep}, the interference is better calculated via the direct integration method to reduce the computing time.

Solution for $[\hat{k}_i]$ exists when a certain kinematic condition is met. The necessary and sufficient condition is
\bea
M_{6l} = \sqrt{(\sum_{i=3}^8 k_{i})^2} > M_{V_1} + M_{V_2} + M_{V_3}.
\label{os_cond_3M}
\eea
An elegant proof for this will be given in \sect{sec:os_cond}. This is called the TPA condition.

The TPA condition generally involves the individual momenta of the final-state neutrinos and charged leptons, with the only exceptional case being the $ZZZ$ process with six 
charged leptons. 
Since the neutrino momenta are unknown in experimental analyses, the TPA condition cannot be imposed in measurements. 
It must however be applied in the simulation of the polarized events.   

In the comparison between theory prediction and data, one must account for off-shell effects which can be large in some kinematic regions 
(see e.g. the region of $P_{T,e\mu\tau} > 200\gev$ in the right plot of \fig{dist:m6l_pT_emt} where the off-shell effect exceeds $20\%$). 
In order to reduce the off-shell effects, suitable analysis cuts should be identified. For example, one can study the correlation between 
the $M_{6l}$ and a transverse mass such as $M_{T,3\ell} = (2P_{T,3\ell}E_{T,\text{miss}}[1-\cos(\Delta\phi(p_{3\ell},p_\text{miss}))])^{1/2}$ 
where $p_\text{miss}$ is the momentum of the three-neutrino system \cite{Dittmaier:2019twg} to find an approximation for the cut in \eq{os_cond_3M}. 
This study has not been attempted in the present work as we would like to explore here a phase space as large as possible.

Some important technical notes are in order. 
For the implementation of \eq{eq:def_TPA}, it is crucial to note that the production and decay amplitudes depend on the on-shell momenta $[\hat{k}_i]$. 
Moreover, as the production and decay amplitudes are calculated using the stable $V_i$ assumption (i.e. infinity-time limits for external states), the resonance decay widths $\Gamma_{V_i}$ 
are neglected in these amplitudes. This means that $\Gamma_{V_i}$ only occur in the prefactor $1/(Q_1Q_2Q_3)$ in \eq{eq:def_TPA}. 
The other decay widths of unstable particles such as the Higgs boson, the top quark or the other massive gauge boson not in the 
triple-resonance system (i.e. the $Z$ boson in the present case) are kept in the calculation of those on-shell amplitudes. 
On the other hand, for the full off-shell calculation of the processes in \eq{eq:process_qq_6l}, the decay widths of all unstable particles, 
including those in the $V_1V_2V_3$ system, are fully taken into account using the complex-mass scheme \cite{Denner:1999gp,Denner:2005fg,Denner:2006ic}. 
Further details of the input parameter scheme for the TPA and off-shell calculations are provided in \sect{sec:res}.
\subsection{The sequential splitting on-shell projection for triboson processes}
\label{sec:map_os}
In this section, we present a general on-shell mapping applicable for any triboson production process. 
Generalization to the case of $n$-resonances with arbitrary decays will be considered in the next section.
The method has the following features: dynamic and democratic. 
It is dynamic in the sense that the mapping depends on the phase-space point. 
This is crucial to obtain a maximal size for the set of on-shell momentum configurations. 
It is democratic because it treats all the vector bosons equally. This is important to preserve the symmetry of the process. 
For example, for the case of the $W^-W^+W^+$ process at hand, a good on-shell mapping should treat the two $W^+$ equally, so that 
the $W^-_XW^+_LW^+_T$ and $W^-_XW^+_TW^+_L$ (with $X$ being any polarization) polarized cross sections are the same.

The on-shell momentum configurations satisfy the following conditions:
\begin{align}
\hat{k}_i^2 &= 0, ~(i=1,..,8), \nonumber \\
\hat{q}_1^2 &= (\hat{k}_3 + \hat{k}_4)^2 = M_{V_1}^2,~\hat{q}_2^2 = (\hat{k}_5 + \hat{k}_6)^2 = M_{V_2}^2,~\hat{q}_3^2 = (\hat{k}_7 + \hat{k}_8)^2 = M_{V_3}^2.
\label{eq:os_cond}
\end{align}

For the initial state particles, we simply choose $\hat{k}_i = k_i$ with $i=1,2$. This leads to the following constraint on $\hat{q}_i$:
\begin{align}
\hat{q}_1 + \hat{q}_2 + \hat{q}_3 = q_1 + q_2 + q_3 = q_{VVV}. 
\end{align} 
The next step is to obtain $\hat{q}_i$. The idea is to split the $VVV$ system into two parts: $A$ and $B$ with $A$ being a single particle while 
$B$ containing two particles. This splitting is obviously not unique. We do not use a fixed splitting, say $A=V_1$ and $B=V_2V_3$, for all 
phase-space points. Rather, the splitting is dynamical and determined point-wise using certain criteria to be specified. 

Assuming that this splitting has been done and we obtain two systems with momenta $q_A$ and $q_B$ in the $VVV$ c.m.f.. 
We then perform a pseudo on-shell mapping for $A$ and $B$, called OSMAP-$AB$, with the conditions
\bea
\hat{q}_A^2 = M_{V_i}^2,\quad \bar{q}_B^2 = m_B^2,
\label{os_cond_AB}
\eea
as done in \cite{Denner:2000bj} for $W^+W^-$ and in \cite{Baglio:2018rcu} for $WZ$. 
Note that $M_{V_i}$ is the pole mass of one of the identified $V_i$ and $m_B^2 = (q_{V_j} + q_{V_k})^2$ is the off-shell invariant mass of the $B$ system. 
The invariant mass of the B system does not change under this mapping. 
The bar notation for the B system is to indicate that this is not yet an on-shell momentum. 
This method is named {\em the Sequential Splitting On-shell Projection} (SSOP).

At this point, a comparison to the Sequential Pairing On-shell Projection (SPOP) used in \cite{Dittmaier:2015bfe,Dittmaier:2019twg} is of benefit to the reader. 
In these works, the on-shell projection is first done on $V_1V_2$ to calculate $\hat{q}_{V_1}$ and $\hat{q}^\prime_{V_2}$, 
then on $V_2V_3$ to calculate $\hat{q}_{V_2}$ and $\hat{q}_{V_3}$. In the second projection, the on-shell projected momentum $\hat{q}^\prime_{V_2}$ is used as an input. 
This makes the on-shell condition for the second mapping complicated when expressing in terms of the original off-shell momenta. 
In our method, as the $q_B^2$ is preserved in the first mapping and the off-shell momenta are input of the second mapping, 
all on-shell conditions can be easily written in terms of the off-shell momenta, 
see \eq{cond_tpa_AB} and \eq{cond_tpa_B_jk}. Numerical 
comparisons between the SPOP and ours will be reported at the end of this section. 

We now continue to describe our method. 
Another key element of our mapping is to fix the spatial direction of $\hat{q}_A$ in the on-shell $AB$ c.m.f. to be the same as that of $q_A$ in the off-shell $AB$ c.m.f.. 
Fixing the angles first is important to make sure that the projected momenta will lie in the physical region. 
Indeed, an important point of the on-shell mapping is this choice of fixing the spatial direction in the {\em center-of-mass frame} (where the momenta 
of the two particles are back-to-back), which we will 
repeatedly employed in the following steps. This idea is not new and has been used in \cite{Denner:2000bj} for the production part and 
in \cite{Denner:2021csi} for the decays. 

The energy component of $\hat{q}_A$ in the on-shell $AB$ c.m.f. is easily calculated using the \eq{os_cond_AB}. 
We have:
\bea
\hat{E}_{A} = \fr{M_{V_i}^2 - m_{B}^2}{2m_{VVV}} + \fr{1}{2} M_{V_i},\quad m_{VVV} = \sqrt{q_{VVV}^2}.  
\eea
After this step, one then obtain $\hat{q}_A$ and $\bar{q}_B$ in the $VVV$ frame with 
\bea
\bar{q}_B = q_{VVV} - \hat{q}_A.
\label{eq_qB}
\eea 
This solution exists when $|\vec{\hat{q}}_A| = (\hat{E}_{A}^2 - M_{V_i}^2)^{1/2}$ is positively real, leading to:
\bea
m_{VVV} > M_{V_i} + m_B.
\label{cond_tpa_AB}
\eea
The next step is to decay $B$ into two on-shell particles $V_j$ and $V_k$. We again employ the OSMAP-$AB$ as follows.  
We first imagine to be in the $\bar{B}$-rest frame and choose the spatial direction of $\hat{q}_{V_j}$ to 
be the same as that of $q_{V_j}$ in the $V_jV_k$ c.m.f.. Specifically, the spatial direction of 
$q_{V_j}$ in the $V_jV_k$ frame (which is different from the $\bar{B}$-rest frame) is first calculated as:
\bea
\vec{n}_{V_j} = \fr{\vec{q}_{V_j}}{|\vec{q}_{V_j}|}.
\eea
Then, in the $\bar{B}$-rest frame we choose the unit vector as:
\bea
\vec{\hat{n}}_{V_j} = \vec{n}_{V_j}.
\label{def_n_Vj}
\eea
The energy component of $\hat{q}_{V_j}$ in the $\bar{B}$-rest frame is simply calculated using the on-shell conditions: 
\bea
\hat{q}_{V_j}^2 = M_{V_j}^2,\quad \hat{q}_{V_k}^2 = M_{V_k}^2.
\eea
We obtain:
\bea
\hat{E}_{V_j} = \fr{M_{V_j}^2 - M_{V_k}^2}{2m_B} + \fr{1}{2} M_{V_j},\quad \hat{E}_{V_k} = m_B - \hat{E}_{V_j}.  
\eea
The magnitude of the spatial momentum reads:
\bea
|\vec{q}_{V_j}| = \sqrt{\hat{E}_{V_j}^2 - M_{V_j}^2}.
\label{def_qq_Vj}
\eea
This solution is real and positive when
\bea
m_B > M_{V_j} + M_{V_k}.
\label{cond_tpa_B_jk}
\eea
\eq{cond_tpa_AB} and \eq{cond_tpa_B_jk} are the necessary and sufficient conditions to have a solution of the on-shell momenta for a given splitting. 
These conditions are stricter than the TPA condition \eq{os_cond_3M}. We will come back to this point at the end of this section.

From \eq{def_n_Vj} and \eq{def_qq_Vj} the on-shell momenta $\hat{q}_{V_j}$ and $\hat{q}_{V_k}$ are known in the $\bar{B}$-rest frame. We then boost them back to 
the $VVV$ frame using the boost vector $\bar{q}_B$ to complete the set of $\hat{q}_{V_i}$, $\hat{q}_{V_j}$, $\hat{q}_{V_k}$. 

Now is the right position to discuss the splitting into $A$ and $B$. The aim is to have a maximal size of the on-shell momentum configurations. 
This means that for each phase-space point under consideration we have to find an $AB$ splitting satisfying the TPA conditions of \eq{cond_tpa_AB} and \eq{cond_tpa_B_jk}. 
In total, there are three possibilities for choosing $A$. If there are more than one possibilities passing the TPA conditions, the one with smallest value of $m_B$ is chosen. 
One can also choose the one with largest $m_B$. We have actually done this and found that both choices give very similar results (for integrated and differential cross sections). 
For integrated cross sections, the differences are within $0.2\%$ for the TTT, TTL, and TLL. For the LLL, the choice of $\text{min}(m_B)$ gives a slightly 
larger cross section, with a difference of $1.3\%$. Since the LLL polarization has a higher priority in polarization studies, the $\text{min}(m_B)$ criterion is adopted in this work.    

The final step is to perform the decays of $\hat{q}_{V_i}$, $\hat{q}_{V_j}$, $\hat{q}_{V_k}$ into leptons. One has to pay attention to 
the lepton assignment here. The simplest way is to re-order the momenta $\hat{q}_{V_i}$, $\hat{q}_{V_j}$, $\hat{q}_{V_k}$ such that 
$\hat{q}_{V_1}$ is placed first, $\hat{q}_{V_2}$ second, and $\hat{q}_{V_3}$ last. One then do the following decays: 
\bea
\hat{q}_{V_1} = \hat{k}_{3} + \hat{k}_{4},\;\; 
\hat{q}_{V_2} = \hat{k}_{5} + \hat{k}_{6},\;\;
\hat{q}_{V_3} = \hat{k}_{7} + \hat{k}_{8},
\eea
using the same idea as above. For example, to find $\hat{k}_{3}$ and $\hat{k}_{4}$, one imagines to be 
in the on-shell c.m.f. of $3$ and $4$ and choose the spatial direction of $\hat{k}_{3}$ to be that 
of $k_{3}$ calculated in the off-shell (3,4) c.m.f.. This is why we have to make the lepton assignment correctly. 
The energy component is calculated as usual using the on-shell conditions. 
Finally, one has to boost the on-shell momenta $\hat{k}_{3}$ and $\hat{k}_{4}$ to the $VVV$ frame using 
the boost vector $\hat{q}_{V_1}$. This idea was used for the decay of $Z\to \ell \ell^\prime$ in \cite{Denner:2021csi}. 
A very similar idea was used in \cite{Dittmaier:2015bfe}, but with an important difference related to the choice of 
the spatial direction. In \cite{Dittmaier:2015bfe} the spatial direction of $\hat{k}_{3}$ is set equal to that of 
$k_{3}$ calculated in the {\em on-shell} (3,4) c.m.f.. Notice that the spatial 
momenta $\vec{k}_{3}$ and $\vec{k}_{4}$ are not back to back in the c.m.f. of $\hat{q}_{V_1}$.

We now report on the numerical comparison with the 
SPOP of \cite{Dittmaier:2015bfe,Dittmaier:2019twg} above mentioned.
We have implemented this procedure with the following improvements. 
The idea is to make the pairing steps dynamical and democratic rather than fixed as done in \cite{Dittmaier:2015bfe,Dittmaier:2019twg} 
to have a maximal size for the set of on-shell momentum configurations. 
Specifically, in the first step, the pair 
$V_iV_j$ is dynamically chosen by checking the condition of $(q_{V_i} + q_{V_j})^2 > (M_{V_i} + M_{V_j})^2$ and choosing 
the pair with smaller value of the invariant mass in case of multiple choices. 
In the second step, the on-shell momentum (of either $V_i$ or $V_j$) chosen to pair with the $q_{V_k}$ 
is the one giving a smaller value of the invariant mass of the resulting pair if both choices pass the condition of \eq{cond_tpa_B_jk}.

We have checked that the numerical results (integrated and differential polarized cross sections) of 
the SSOP and those of the 
SPOP are very close. 
For the integrated cross sections, differences are within $0.1\%$ for the $WWW$ processes. 
The SSOP is chosen in this work because it is simpler and more elegant. 

Additionally, \bib{Dittmaier:2019twg} introduced another on-shell mapping for triboson production 
where all three on-shell momenta of the three resonances are simultaneously calculated. 
We have however not yet performed any comparison with this 
as we do not see any advantage of this mapping.

Finally, during the revision of this manuscript, we became aware of a general on-shell projection described 
in \cite{Denner:2024xul} which has been implemented in the computer program MoCaNLO \cite{Denner:2026phn}. 
The method of \cite{Denner:2024xul} and the SSOP share many features, in 
particular the key feature that an off-shell invariant mass is unchanged under a projection (see \eq{os_cond_AB}). 
Additionally, \bib{Denner:2024xul} introduces a new Simultaneous On-shell 
Projection (SOP) to calculate simultaneously all on-shell momenta of $n$ resonances or decay products.
\subsection{On-shell projection for $n$-resonance processes} 
The SSOP above described 
can be easily extended to the case of $n$ resonances. In the first step, one particle 
is on-shell projected and the remaining $(n-1)$ off-shell particles form the system $B_1$ by using the 
OSMAP-$AB$. In the second step, the second on-shell momentum is calculated by performing the same on-shell projection on the $B_1$ set. 
The remaining off-shell momenta form the new set $B_2$. 
Thus, after $(n-1)$ steps, all $n$ on-shell momenta $\hat{q}_{V_i}$ are calculated. 
As for the case of triple resonances, the splitting into $A$ and $B$ at each step should be done dynamically and democratically 
rather than using a fixed ordering. 

Finally, the on-shell momenta of the decay products can be obtained 
by using the same sequential splitting method. 
Specifically, we consider the decay $q_i \to k_1 + k_2 + \ldots + k_m$ of the resonance $V_i$ with $k_j^2 = m_j^2$ ($j=\overline{1,m}$). 
In the first step, the OSMAP-$AB$ is applied to the set of $m$ decay products. 
As the on-shell mapping conditions of 
$M_{V_i} > \sum_j m_j$ and 
$(k_{j_1} + \ldots + k_{j_l})^2 > (m_{j_1} + \ldots + m_{j_l})^2$ with $2\le l \le m$ 
are all satisfied here, 
the simplest choice is splitting the set of $m$ decay particles into two sets $C_1$ and $D_1$, where $C_1 = \{k_1\}$ and $D_1 = \{k_2, \ldots, k_m\}$, 
keeping the original ordering of the decay products. 
The on-shell momentum $\hat{k}_1$ and the new off-shell momentum $\bar{q}_{D_1}$ are then calculated as above described (see \eq{os_cond_AB} and \eq{eq_qB}).
In the second step, the same procedure is applied to the set $D_1$ to calculate $\hat{k}_2$. 
Thus, after $(m-1)$ steps, all $m$ on-shell momenta of 
the decay products of the resonance $V_i$ are calculated. 

For the case of nested resonances, e.g. $q_i \to R (k_1 k_2 k_3) k_4 \ldots k_m$ where $R$ is a 
nested resonance with mass $M_{R}$, we can enforce an extra on-shell condition of $(\hat{k}_1 + \hat{k}_2 + \hat{k}_3)^2 = M_{R}^2$ easily 
by choosing $A = \{k_1, k_2, k_3\}$, $B = \{k_4, \ldots, k_m\}$ with $m_A^2 = M_R^2$ and $m_B^2 = (k_4 + \ldots + k_m)^2$ and 
apply the OSMAP-$AB$ on the set $\{k_1, \ldots, k_m\}$ to find the on-shell momentum $\hat{q}_A$ and the new momentum $\bar{q}_B$. 
The subsequent decays of $A$ and $B$ then follow as above described. As the enforcement of an intermediate on-shell condition can be done at any step of the 
sequential splitting procedure, the number of nested resonances can be arbitrary. 
We used this procedure in \cite{Dao:2024ffg} for the case of top-quark decay with 
a nested $W$ resonance.

Finally, a comparison between the SOP \cite{Denner:2024xul} and the SSOP is noted. While the SSOP is fully analytical, 
the SOP is a numerical method for the case of $n\ge 3$. 
The SOP involves a scaling factor $\alpha$, see \eq{eq_alpha} and the text there. 
Already for the case of $n=3$, this equation is rather complicated and no analytical solution of $\alpha$ is available. 
A numerical solution satisfying the required precision level of on-shellness must be implemented as done e.g. in MoCaNLO \cite{Denner:2026phn}.  
\subsection{The necessary and sufficient condition for on-shell momenta} 
\label{sec:os_cond}
We provide here a simple proof of the 
necessary and sufficient condition to have a solution of the on-shell momenta for the general process of $n$ resonance production. 

We consider this process
\bea
p_1 + p_2 \to k_1 + k_2 + \ldots + k_n,
\eea
where $k_i$ are off-shell momenta, i.e. $k_i^2 \neq M_i^2$ with $M_i$ being the pole mass of the $i$-th resonance. 
The question is: What is the necessary and sufficient condition to have a solution for $\hat{k}_i$ satisfying the following conditions:
\begin{align}
\sum_{i=1}^n \hat{k}_i &= Q,\label{os_cond_gen1}\\
\hat{k}_i^2 &= M_i^2,\label{os_cond_gen2} 
\end{align}
where $Q = p_1 + p_2$. 

The answer is
\bea
\sqrt{Q^2} > \sum_{i=1}^{n} M_i. \label{os_cond_gen}
\eea

If one uses a SSOP, then there are $(n-1)$ conditions, one condition for each splitting 
(see \eq{cond_tpa_AB} and \eq{cond_tpa_B_jk} for the case $n=3$). Choosing a different splitting order gives 
another set of conditions. The overall condition of each set is always stricter than \eq{os_cond_gen}. 
It is possible that if all permutations of splittings are considered, the result of 
\eq{os_cond_gen} can be proven. This path is not easy, hence we will not follow it.

It turns out that there is a very elegant way to prove \eq{os_cond_gen} using the SOP. 
The argument runs as follows. First, from \eq{os_cond_gen1} and \eq{os_cond_gen2} one then has 
the condition \eq{os_cond_gen} (the best way to see this is going to the rest frame of $Q$). 
To prove this is sufficient, one must prove that at least a solution for $\hat{k}_i$ exists. 
This is indeed the case and a solution can be found as follows. 

First, we boost all the off-shell momenta $k_i$ to the rest frame of $Q$. In this frame, we have $\sum_{i} \vec{k}_i = 0$. 
Then rescale all the spatial vectors $\vec{k}_i$ by a common factor $\alpha > 0$ to obtain $\vec{r}_i = \alpha\vec{k}_i$. 
We then choose $\hat{k}_i = (\sqrt{M_i^2 + |\vec{r}_i|^2},\vec{r}_i)$ satisfying \eq{os_cond_gen2}, as done in the SOP \cite{Denner:2024xul}.
For this solution to satisfy \eq{os_cond_gen1} we must have 
\bea
f(t) = \sum_{i=1}^{n} \sqrt{M_i^2 + t |\vec{k}_i|^2} = \sqrt{Q^2},\;\;\; t = \alpha^2.
\label{eq_alpha}
\eea
As $f(0)=\sum_i M_i$ and $f(t)$ is {\em strictly increasing}, there is always a {\em unique} solution for $\alpha$ when $\sqrt{Q^2} > \sum_i M_i$. 
This completes our proof. 

It is interesting to note that though the on-shell projection for multi-resonance production processes has been carefully studied in 
\cite{Dittmaier:2015bfe,Dittmaier:2019twg,Denner:2024xul}, the necessary and sufficient condition for the existence 
of the on-shell momenta has never been written down. At first sight, one suspects that the condition 
\eq{os_cond_gen} is enough. However, to prove that it is sufficient for the case of $n \ge 3$ is nontrivial. 
It is possible for us to write down a complete proof here thanks to the new simultaneous mapping of \cite{Denner:2024xul}, which 
we became aware of during the revision of this paper.
\subsection{Code implementation}
\label{sec:code}
The calculation of this work has been implemented in our private computer program MulBos (for MultiBoson production). 
This program was first created for calculating polarized cross sections of diboson processes. 
Results at the NLO QCD+EW level for $WZ$, $W^+W^-$, and $ZZ$ have been published in \cite{Le:2022ppa,Dao:2023kwc,Carrivale:2025mjy}. 
This work marks the first extension of MulBos to the class of triboson processes. 

The helicity amplitudes for the production and decay processes are generated by FeynArts \cite{Hahn:2000kx} and FormCalc \cite{Hahn:1998yk}. 
Phase-space integration is done using the Vegas Monte-Carlo method \cite{Lepage:1977sw} as implemented in the computer program BASES \cite{Kawabata:1995th}. 
Additional importance sampling is implemented in the phase-space generation step to smooth out the Breit-Wigner distributions of the 
resonances. This is done with the help of resonance mapping routines publicly available in the open-source code VBFNLO-3.0 \cite{Baglio:2024gyp}. 
 
For cross-checking purpose, we have performed two independent implementations. In one implementation, FeynArts-3.4 and FormCalc-6.0 
were used to generate the helicity amplitudes. In the other, FeynArts-3.11 and FormCalc-9.10 were employed. Phase-space generation and on-shell 
mapping have been independently implemented. Results of the two codes agree 
within the statistical errors. Furthermore, for the full off-shell cross sections, good agreement with VBFNLO-3.0 has been found.
\section{Numerical results}
\label{sec:res}
\subsection{Input parameters}
\label{sec:input}
For the numerical results presented in this section, the proton-proton center-of-mass energy is $\sqrt{s} = 13.6$ TeV. 
As default, the factorization scale is chosen dynamically as $\mu = M_{6l}/2$. 
For the parton distribution functions (PDF), the set \texttt{NNPDF40\_lo\_as\_01180} \cite{NNPDF:2021njg} is used via the 
LHAPDF6 library \cite{Buckley:2014ana}.

Model parameters are $G_F = 1.1663788 \times 10^{-5}~\text{GeV}^{-2}$ (Fermi constant) and 
$M_W = 80.3692~\text{GeV}$, $M_Z = 91.1880~\text{GeV}$ according to the latest data from Particle Data Group \cite{ParticleDataGroup:2024cfk}. 
Other relevant parameters include $\Gamma_W = 2.14~\text{GeV}$, $\Gamma_Z = 2.4955~\text{GeV}$, 
$M_H = 125.20 ~\text{GeV}$, $\Gamma_H = 0.0037~\text{GeV}$. The leptons and quarks are approximated as massless. 
The top quark does not occur in this calculation. 

For the TPA calculations, the electromagnetic coupling constant is obtained via the Fermi constant using 
$\alpha_\text{em} = \sqrt{2} G_F M_W^2 (1 - M_W^2/M_Z^2)/\pi$. This gives $1/\alpha_\text{em} \approx 132.10015$ and $\sin^2\theta_W = 1-M_W^2/M_Z^2 \approx 0.22321$ with 
$\theta_W$ being the weak-mixing angle. 
For the full off-shell calculation (named full for short), the complex-mass scheme is used. At LO, this is simply done as follows. 
In the first step, the pole mass and total decay width of the $W$ and $Z$ bosons are shifted as (with $V=W,Z$):
\bea
\hat{M}_V = \fr{M_V}{\sqrt{1+(\Gamma_V/M_V)^2}},\quad 
\hat{\Gamma}_V = \fr{\Gamma_V}{\sqrt{1+(\Gamma_V/M_V)^2}}.
\eea
Then, the idea is to replace $M_V^2$ with the complex mass $\tilde{M}_V^2 = \hat{M}_V^2 - i\hat{M}_V\hat{\Gamma}_V$ everywhere 
in the helicity amplitudes keeping the electromagnetic coupling constant $e = \sqrt{4\pi\alpha_\text{em}}$ as an independent real parameter. 
In particular the $\sin\theta_W$ and $\cos\theta_W$ are made complex with $\cos^2\theta_W = \tilde{M}_W^2/\tilde{M}_Z^2$. 
The value of $\alpha_\text{em}$ is calculated as $\alpha_\text{em} = \sqrt{2} G_F \hat{M}_W^2 (1 - \hat{M}_W^2/\hat{M}_Z^2)/\pi$. 
Numerically, the new values of $\alpha_\text{em}$ and $\sin^2\theta_W$ in the complex-mass scheme then read 
$1/\alpha_\text{em} \approx 132.21216$ and $\sin^2\theta_W \approx 0.22319-5.73970\times 10^{-4}i$,
which are slightly shifted compared to the ones of the TPA.

Since our purpose is to survey the magnitudes of polarized cross sections across the whole phase space, focusing on the triple resonance region, 
a simple set of kinematic cuts has been used. They are:
\begin{align}
& P_{T,\ell} > 15 ~\text{GeV},~ \abs{\eta_\ell} \le 2.7,~ \Delta R (\ell,\ell^\prime) > 0.1, \crn
& m_{W^{+}W^-}> 130~\text{GeV}, \label{eq:kin_cuts}
\end{align}
where $\ell$ is a charged lepton (i.e. $e$, $\mu$ or $\tau$), and $\Delta R (\ell,\ell^\prime) = \sqrt{(\eta_{\ell}-\eta_{\ell^\prime})^2 + (\phi_{\ell}-\phi_{\ell^\prime})^2}$ 
with $\phi_\ell$ and $\eta_\ell$ being the azimuthal angle and pseudo-rapidity of the lepton $\ell$. These cuts are implemented in the proton-proton c.m.f. (Lab frame). 
The $W^+W^-$ invariant mass cut is to remove the $H\to 4~\text{leptons}$ resonance which is of no interest in this study as 
we are focusing on the triple-resonance region. This cut has an impact in the full off-shell case. 
For the TPA calculations its effects are negligible due to the TPA condition. 

Before presenting our findings, we stress again here that the polarized cross sections are calculated in the $WWW$ c.m.f..
\subsection{Integrated cross sections}
\label{sec:xs_int}
\begin{table}[h!]
	\centering 
	\renewcommand{\arraystretch}{1.3}	
	\begin{bigcenter}	 
		{\fontsize{9.0}{9.0}
				\begin{tabular}{|c|c|c|c|c|c|} \hline
					& {\fontsize{9.0}{9.0} $\sigma_{\text{fix}}$ $[\text{ab}]$} & $f_{\text{fix}}$ [$\%$] & 
					 $\sigma_{\text{dyn}}$ $[\text{ab}]$ & $f_{\text{dyn}}$ [$\%$] & $\delta_\sigma$ [$\%$] \\
					\hline
					{\fontsize{9.0}{9.0}$\text{Full off-shell}$}  & $122.29(2)^{+0.06\%}_{-0.46\%}$ & -- &
					 $120.13(2)^{+0.00\%}_{-0.28\%}$ & -- & $+1.8$ \\
					{\fontsize{9.0}{9.0}$\text{Unpol. TPA}$}  & $118.38(2)^{+0.15\%}_{-0.60\%}$ & $100$ & 
					 $116.52(2)^{+0.00\%}_{-0.37\%}$ & $100$ & $+1.6$ \\
					\hline
					{\fontsize{9.0}{9.0}$\text{TTT}$} & $60.75(1)^{+0.00\%}_{-0.48\%}$ & $51.3^{+0.46\%}_{-0.63\%}$ & 
					 $59.30(1)^{+0.05\%}_{-0.65\%}$ & $50.9^{+0.42\%}_{-0.50\%}$ & $+2.4$ \\
					{\fontsize{9.0}{9.0}$\text{TTL}$} & $43.28(1)^{+0.38\%}_{-1.09\%}$ & $36.6^{+0.23\%}_{-0.49\%}$ & 
					 $42.83(1)^{+0.17\%}_{-0.72\%}$ & $36.8^{+0.33\%}_{-0.35\%}$ & $+1.1$ \\
					{\fontsize{9.0}{9.0}$\text{TLL}$} & $10.581(2)^{+1.26\%}_{-2.16\%}$ & $8.9^{+1.10\%}_{-1.57\%}$ & 
					 $10.578(2)^{+0.97\%}_{-1.82\%}$ & $9.1^{+1.13\%}_{-1.45\%}$ & $0.0$ \\
					{\fontsize{9.0}{9.0}$\text{LLL}$} & $1.5998(4)^{+1.41\%}_{-2.27\%}$ & $1.4^{+1.25\%}_{-1.69\%}$ &  
					 $1.5982(4)^{+1.01\%}_{-1.93\%}$ & $1.4^{+1.16\%}_{-1.56\%}$ & $+0.1$ \\
					{\fontsize{9.0}{9.0}$\text{Interference}$}  & $2.176(7)^{+1.66\%}_{-2.16\%}$ & $1.8^{+1.51\%}_{-1.58\%}$ & 
					 $2.182(7)^{+1.39\%}_{-1.63\%}$ & $1.9^{+1.54\%}_{-1.26\%}$ & $-0.3$ \\
					\hline	
				\end{tabular}
		}
		\caption{Unpolarized and triply-polarized cross sections in attobarn (ab), together with the corresponding polarization fractions, calculated at LO using the TPA approach in the $WWW$ c.m.f. for the process $pp \rightarrow W^- W^{+} W^{+} \rightarrow e^- \bar{\nu}_e \mu^+ \nu_\mu \tau^+ \nu_\tau + X$ using the central fixed and dynamical factorization scales. 
		The corresponding full off-shell unpolarized cross sections are also provided.
		Statistical uncertainties are given in parentheses while scale uncertainties are provided in percentage as superscripts and subscripts. 
		The scale uncertainties of the polarizarion fractions are calculated by using the correlated-scale variation.
		In the last column, the relative difference between the fixed and the dynamical scales is given for the cross sections.}
		\label{tab:mpp_mu0}
	\end{bigcenter}	
\end{table}
\begin{figure}[h!]
	\centering
	\includegraphics[width=0.49\textwidth]{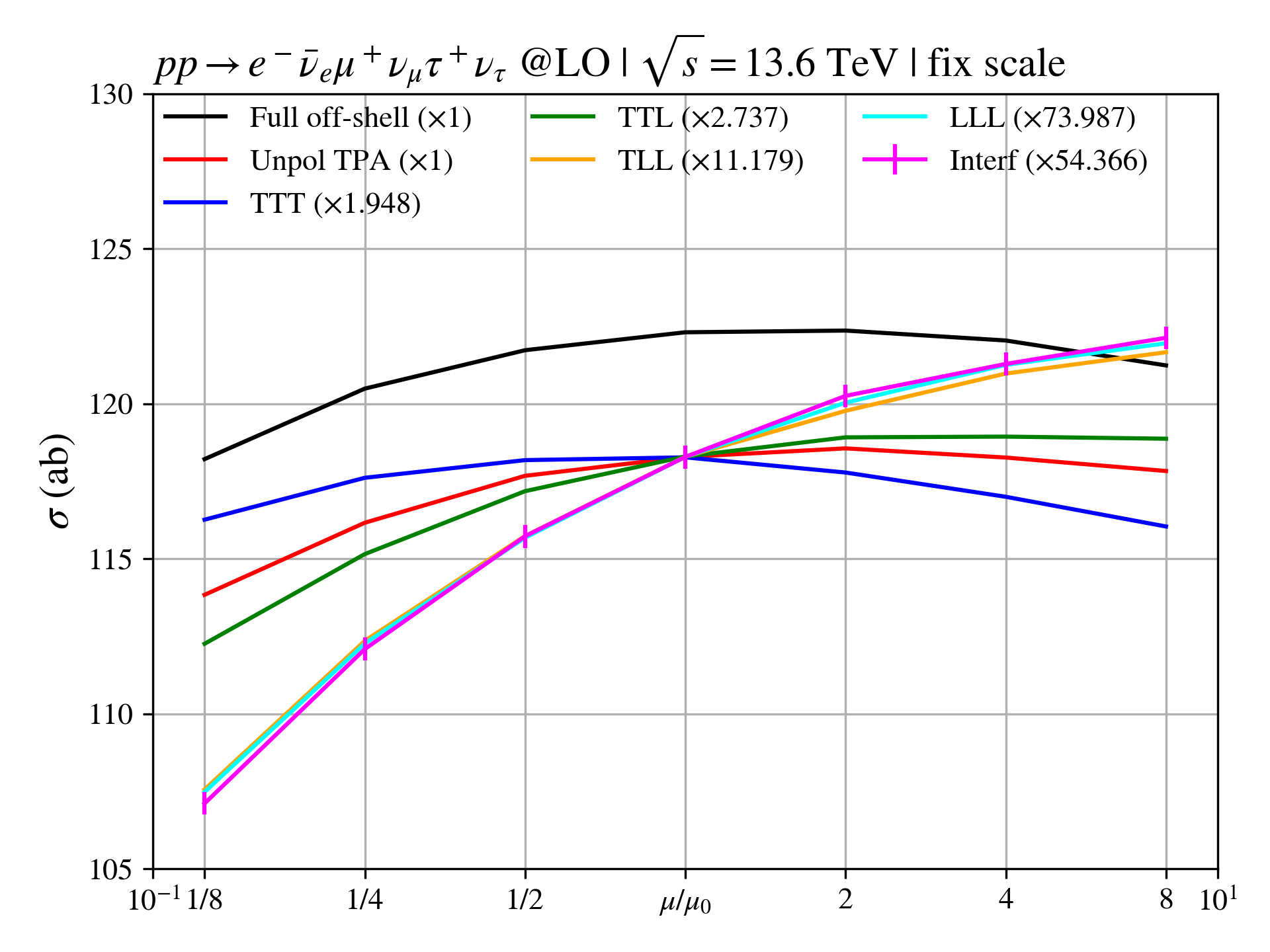}
	\includegraphics[width=0.49\textwidth]{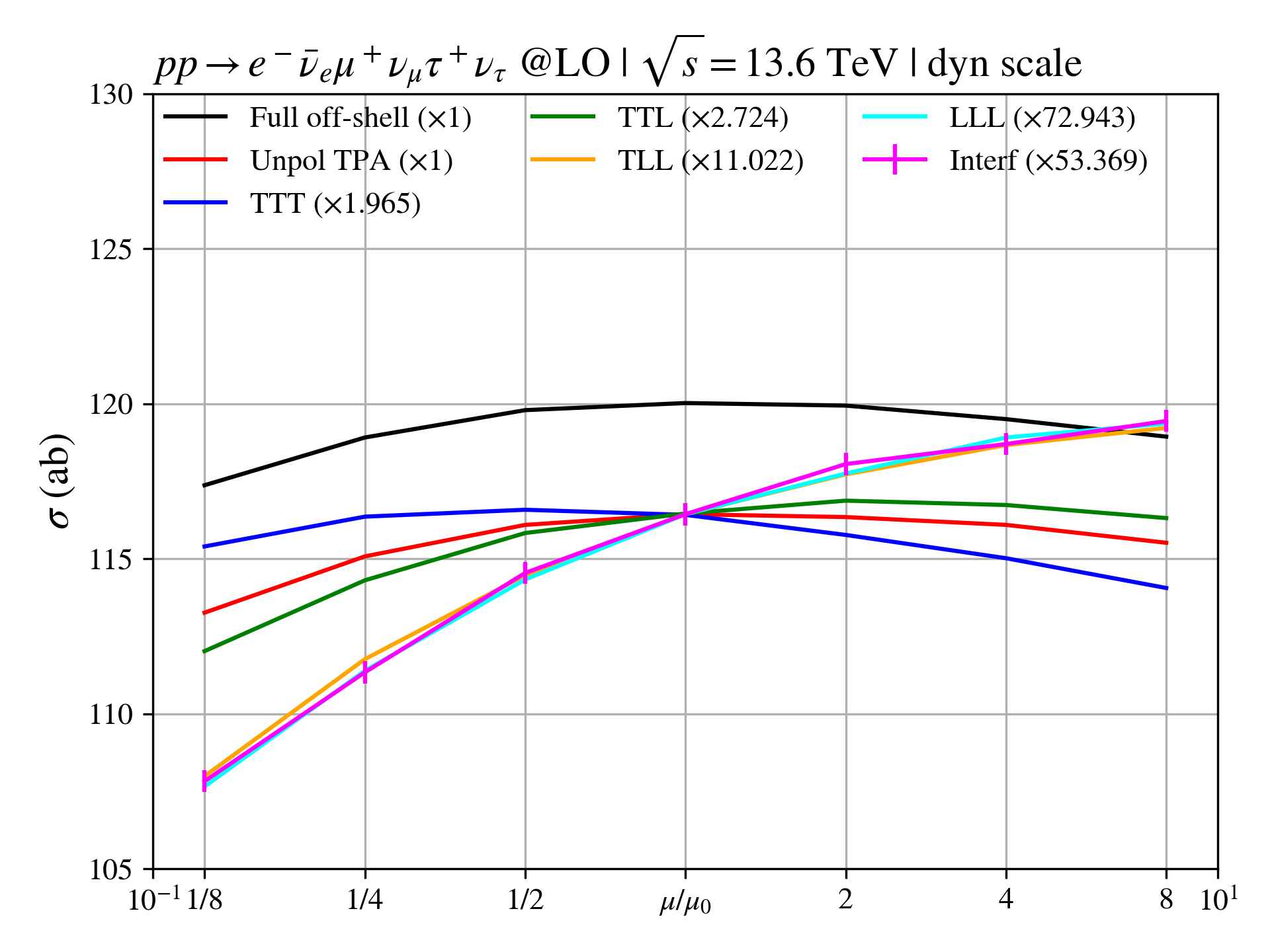} 
	\caption{Cross sections as functions of the factorization scale parameter $\xi = \mu/\mu_0$ for fixed (left) and dynamical (right) scales. 
	The vertical bars on the interference curve represent the statistical errors.} 
	\label{fig:xi_dep}
\end{figure}
We first present results for the integrated cross sections. As the dependence on the factorization scale is an important issue for the estimation of theoretical uncertainties, 
we make a comparison here 
between two choices: fixed scale choice of $\mu_\text{fix} = \xi (3M_W)/2$ and the default dynamical scale $\mu_\text{dyn} = \xi M_{6l}/2$. 
Results for the case of $\xi = 1$ are provided in \tab{tab:mpp_mu0}, while the $\xi$ dependence is plotted in \fig{fig:xi_dep}. 

Besides the statistical uncertainties, the scale uncertainties are also provided in \tab{tab:mpp_mu0}. 
They are calculated by deviating the scale from its central position ($\mu_0$) by a factor of two, up and down. 
This gives asymmetric errors. The usual polarization fractions, calculated with respect to the unpolarized TPA result, are also given. 
The relative difference between the fixed-scale cross section and dynamical-scale one is calculated as $\delta_\sigma = (\sigma_\text{fix}-\sigma_\text{dyn})/\sigma_\text{dyn}$.

We first observe that, for the cross sections, the difference between the fixed scale and the dynamical scale is less than $3\%$, in magnitude, 
for all cases. The scale uncertainties reflect this difference in the TTL, TLL, LLL, and interference cases. For the remaining cases 
of unpolarized and TTT cross sections, the scale uncertainties are too small. This can be understood by looking at \fig{fig:xi_dep}. 
The central scale $\mu = \mu_0$ is near the maximum position for the unpolarized and TTT cases, leading to very small scale uncertainties. 
In \bib{Dittmaier:2019twg}, larger scale uncertainties are reported. This is mainly due to the different kinematic cuts.

It is however not only the position of the central scale that is important but also the shape of the scale dependence. 
The slope increases with the addition of longitudinal mode. There are big changes going from TTT to TTL and to TLL, 
while the change is significantly reduced from TLL to LLL. 

It is interesting to note the size of the interference fraction, about $2\%$ for both scale choices. 
Using the SPOP or the SSOP with the largest $m_B$ (see \sect{sec:map_os}) gives the same result. 
For the process $W^+W^-W^-$, it is slightly smaller, of $1.6\%$, as shown in \appen{appen_pmm}. 
The scale uncertainties of the interference cross section are around $2\%$ for both fixed and dynamical scales, 
being of the same size as those of the LLL and TLL polarizations.   
These values are larger than the difference between the fixed-scale and dynamical-scale cross sections. 
These scale uncertainties can be better understood by looking at the scale dependence in \fig{fig:xi_dep}. 
We see that the $\mu$ dependence of the interference contribution is very similar to the LLL one. 
For the $W^+W^-W^-$ process, there is a clearer difference between them, as shown in \fig{fig:xi_dep_pmm}, with 
the interference term increasing faster with the scale. 

We now discuss the issue of scale choice. For integrated cross sections, it seems that both fixed and dynamical scales  
are good choices because the dominant contribution comes from the triple-resonance threshold region. 
This argument is also valid for diboson $VV^\prime$ productions. 
Here the fixed scale of $\mu_0 = (M_V + M_{V^\prime})/2$ is commonly used because it gives 
larger cross sections which agree better with the measurement values. 
For differential cross sections, it has been widely recognized that dynamical scale is a better choice, in particular 
for transverse momentum or invariant mass distributions at high-energy regions. 

The scale dependence shown in \fig{fig:xi_dep} gives us a new view on this issue. 
While for the unpolarized and TTT cross sections both choices gives the same dependence across the range of $\mu/\mu_0 \in [1/8,8]$, 
the factorization scale dependence is significantly improved when using the dynamical scale for the TTL, TLL, LLL, and interference 
cross sections. We will therefore use the dynamical scale as the default choice for both integrated and differential cross sections.
  
Finally, we close this section by noting the smallness of the LLL fraction, being $1.4\%$ and smaller than the interference term. 
With increasing number of longitudinal modes, 
the ratios $\sigma_\text{TTL}/\sigma_\text{TTT}$, $\sigma_\text{TLL}/\sigma_\text{TTL}$, and $\sigma_\text{LLL}/\sigma_\text{TTL}$ are $0.72$, $0.25$, and $0.15$ 
for the dynamical scale, respectively. 
These values are essentially the same for the fixed scale case or for the $W^+W^-W^-$ process (see \appen{appen_pmm} for precise numbers).
For comparison, the LO ratios $\sigma_\text{TL}/\sigma_\text{TT}$ and $\sigma_\text{LL}/\sigma_\text{TL}$ (TL here means TL+LT) are 
$0.31$ and $0.30$ for $W^+W^-$ \cite{Denner:2023ehn,Dao:2023kwc}, 
$0.29$ and $0.38$ for $W^+Z$ \cite{Le:2022lrp}, 
$0.32$ and $0.38$ for $W^-Z$ \cite{Le:2022ppa}, 
$0.34$ and $0.25$ for $ZZ$ \cite{Carrivale:2025mjy}. 

As QCD corrections in the triboson and diboson processes share many common features (e.g. initial-state radiation, virtual corrections), 
it is worth summarizing here the higher-order results available for the diboson case. 
We are interested in the change of the polarization fractions calculated in the $VV$ c.m.f. when going from LO to (N)NLO QCD. 
NLO EW corrections are sometimes included for the sake of accuracy, but they are negligible when considering integrated cross sections.
For the $W^+W^-$ process without jet veto and excluding the bottom-quark contribution, the $f_\text{LL}$ changes from $6.6\%$ \cite{Dao:2023kwc} at LO to $5.7\%$ at NLO QCD \cite{Dao:2024ffg}. 
For the $W^+Z$ process \cite{Pelliccioli:2025com}, it goes from $7.9\%$ at LO to $5.7\%$ at NLO QCD+EW, and to $5.6\%$ at NNLO QCD + NLO EW. 
For the $W^-Z$ process \cite{Le:2022ppa}, it is $8.6\%$ at LO and $5.9\%$ at NLO QCD. 
For the $ZZ$ case \cite{Carrivale:2025mjy}, it varies from $5.8\%$ at LO to $5.9\%$ at NLO QCD, and to $6.1\%$ at NNLO QCD. 
NNLO QCD results for the $W^+W^-$ are provided in \cite{Poncelet:2021jmj} but not usable here because 
they are calculated in the Lab frame. 
One thus sees that higher-order effects are mildest for the $ZZ$ case where they increase the LL fractions slightly. 
For the remaining cases involving the $W$ boson the effects are significantly stronger, but negative.
The largest reduction is $31\%$ when going from LO to NLO QCD for the $W^-Z$ process, and the smallest reduction 
is $14\%$ for the $W^+W^-$ case. 
The same survey for the TL fraction gives positive corrections, ranging from $+7\%$ for $ZZ$, $+20\%$ for $W^+W^-$, 
$+41\%$ for $W^-Z$, $+48\%$ for $W^+Z$ when going from LO to NLO QCD. NNLO QCD corrections are much smaller.  
From these results, we see that higher-order corrections affect different polarization fractions in different ways. 
It is therefore possible that the LLL fraction would get a positive correction from higher-order effects. 
However, this enhancement is not expected to bring the LLL fraction to the level of tens of percent. 
It is still very challenging to measure the $\sigma_\text{LLL}$ at the LHC even in this optimistic scenario. 

The smallness of the LLL fraction should not discourage us from performing its measurement as an upper limit 
will help us put new constraints on new-physics parameters such as the anomalous gauge couplings.
\subsection{Kinematic distributions}
\label{sec:dist}
\begin{figure}[h!]
	\centering
		\includegraphics[width=0.49\textwidth]{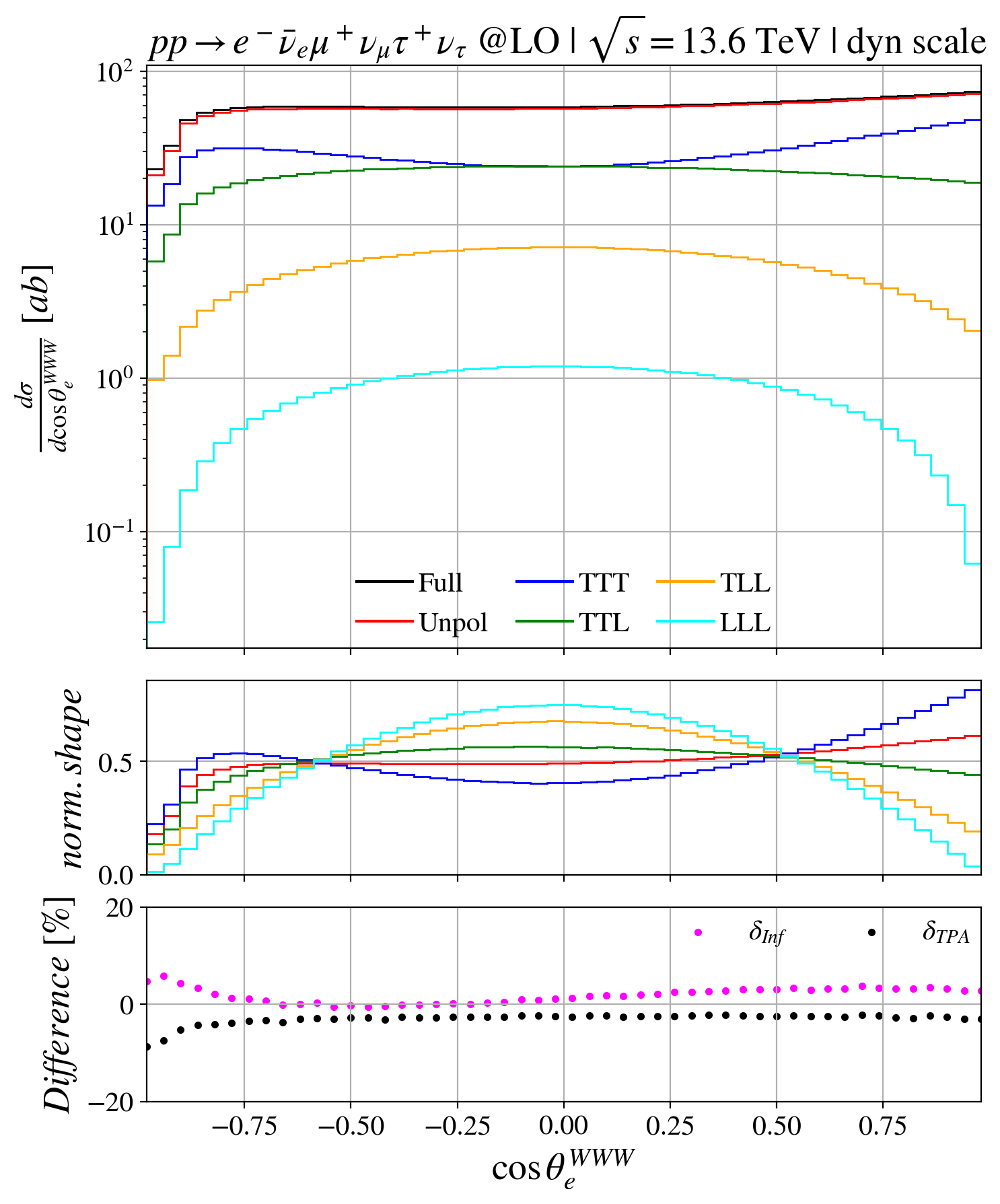}
		\includegraphics[width=0.49\textwidth]{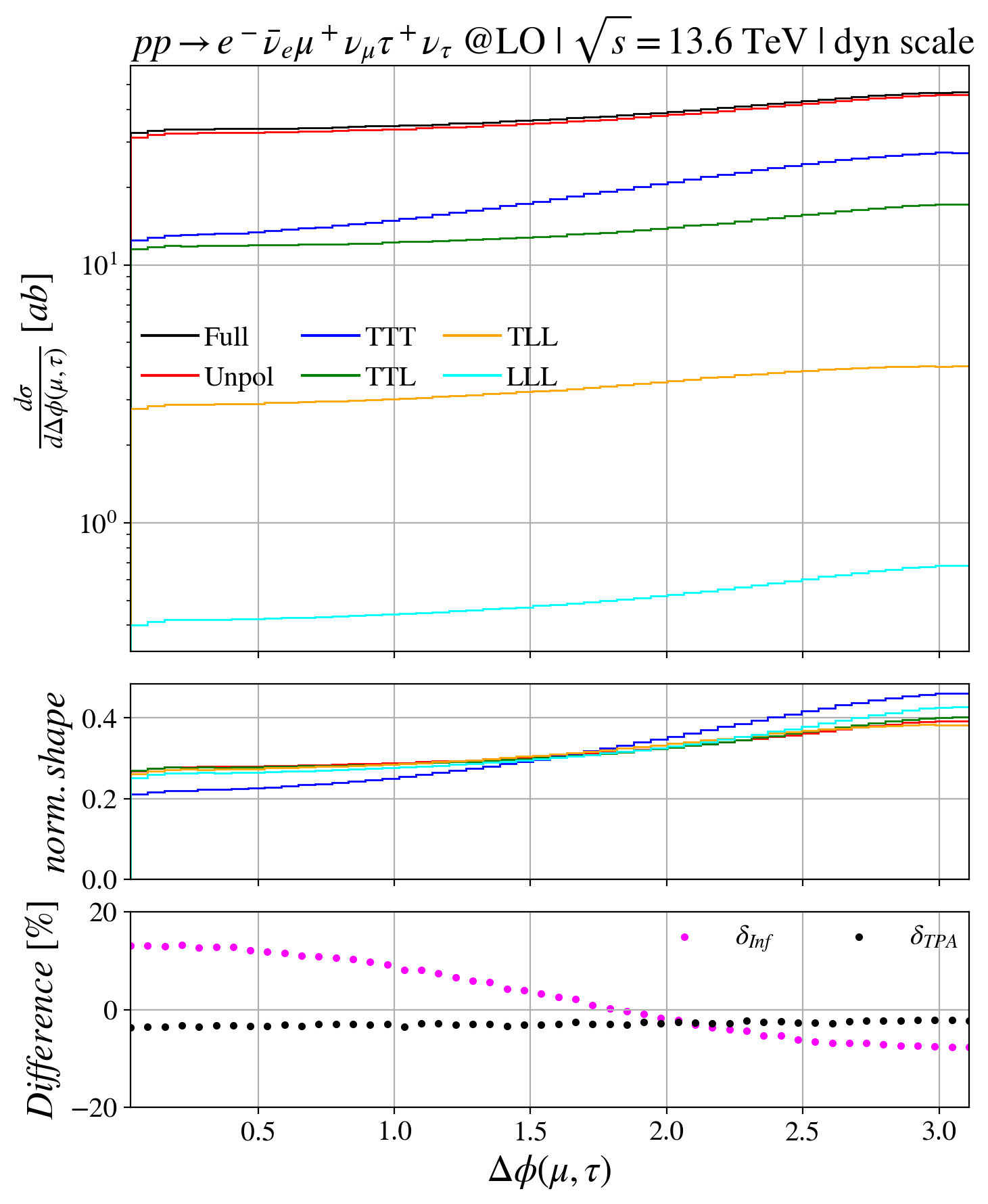} 
	\caption{Distributions in $\cos \theta_e^{WWW}$ (left) and in azimuthal-angle difference between 
	the muon and the tauon (right) are shown. The polar angle $\theta_e^{WWW}$ is defined in the $WWW$ c.m.f. (see the text for details). 
	The top panels show the values of the LO differential cross sections. The middle panels show the normalized distributions (areas are equal to unity), while the bottom panels display the   interference contribution and the off-shell effect (difference between the unpolarized TPA and the full off-shell).} 
	\label{dist:thetaE_DeltaPhi_MT}
\end{figure}
\begin{figure}[h!]
	\centering
		\includegraphics[width=0.49\textwidth]{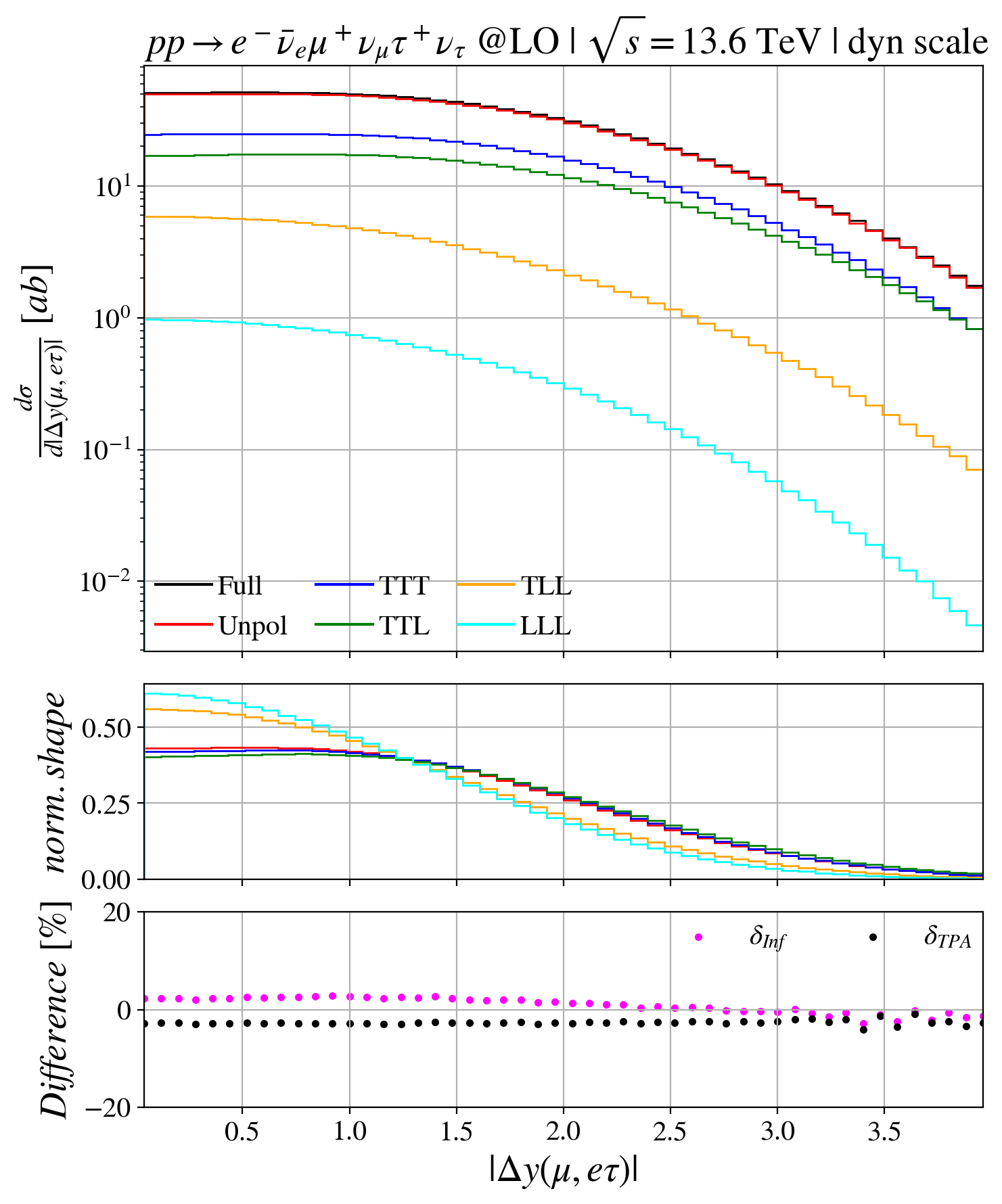} 
		\includegraphics[width=0.49\textwidth]{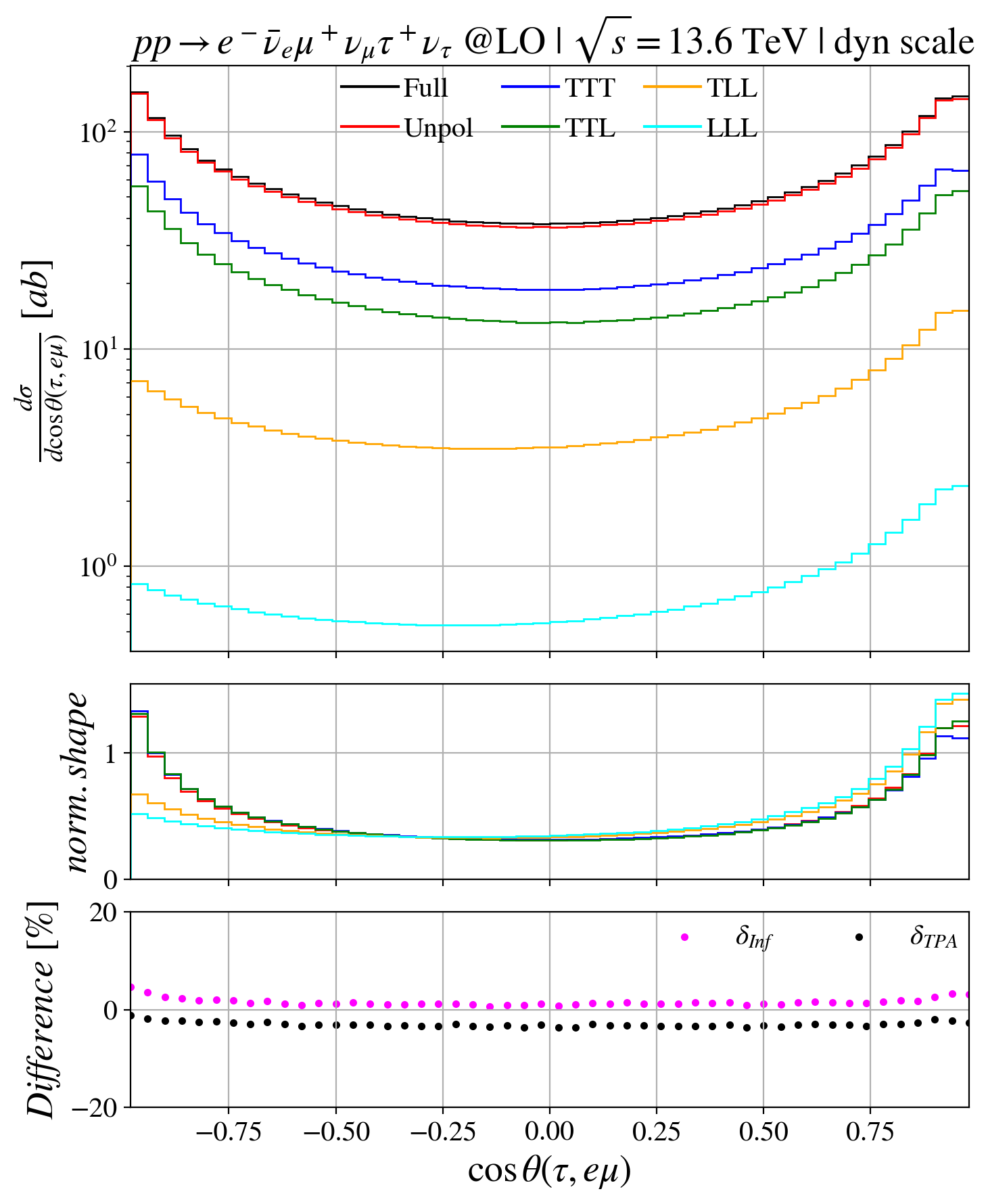}
	\caption{Same as \fig{dist:thetaE_DeltaPhi_MT} but for the rapidity difference between the muon and the electron-tauon system (left) and cosine of the angle 
	between the tauon and the electron-muon system (right).} 
	\label{dist:y_theta_emt}
\end{figure}
\begin{figure}[h!]
	\centering
		\includegraphics[width=0.49\textwidth]{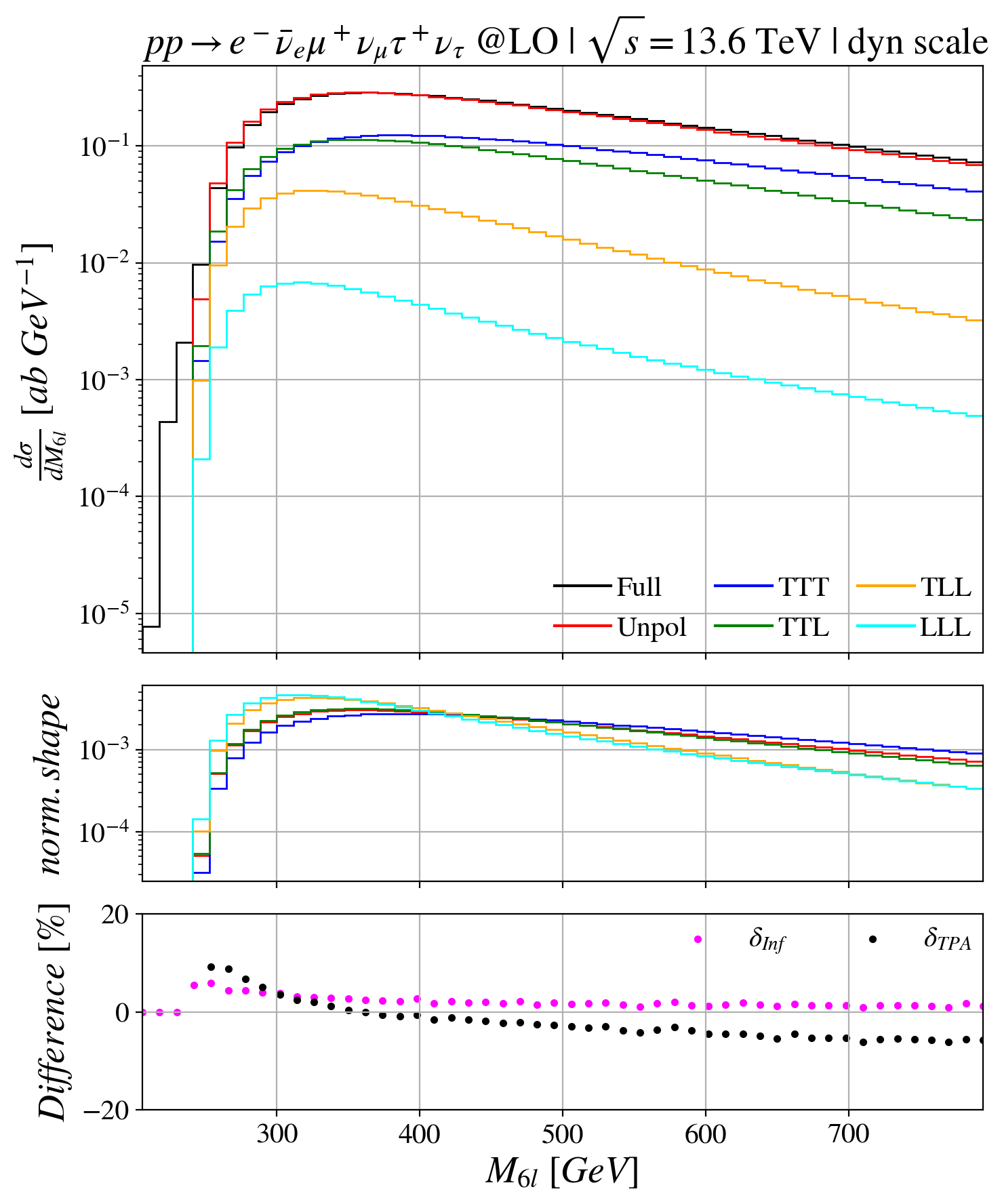}
		\includegraphics[width=0.49\textwidth]{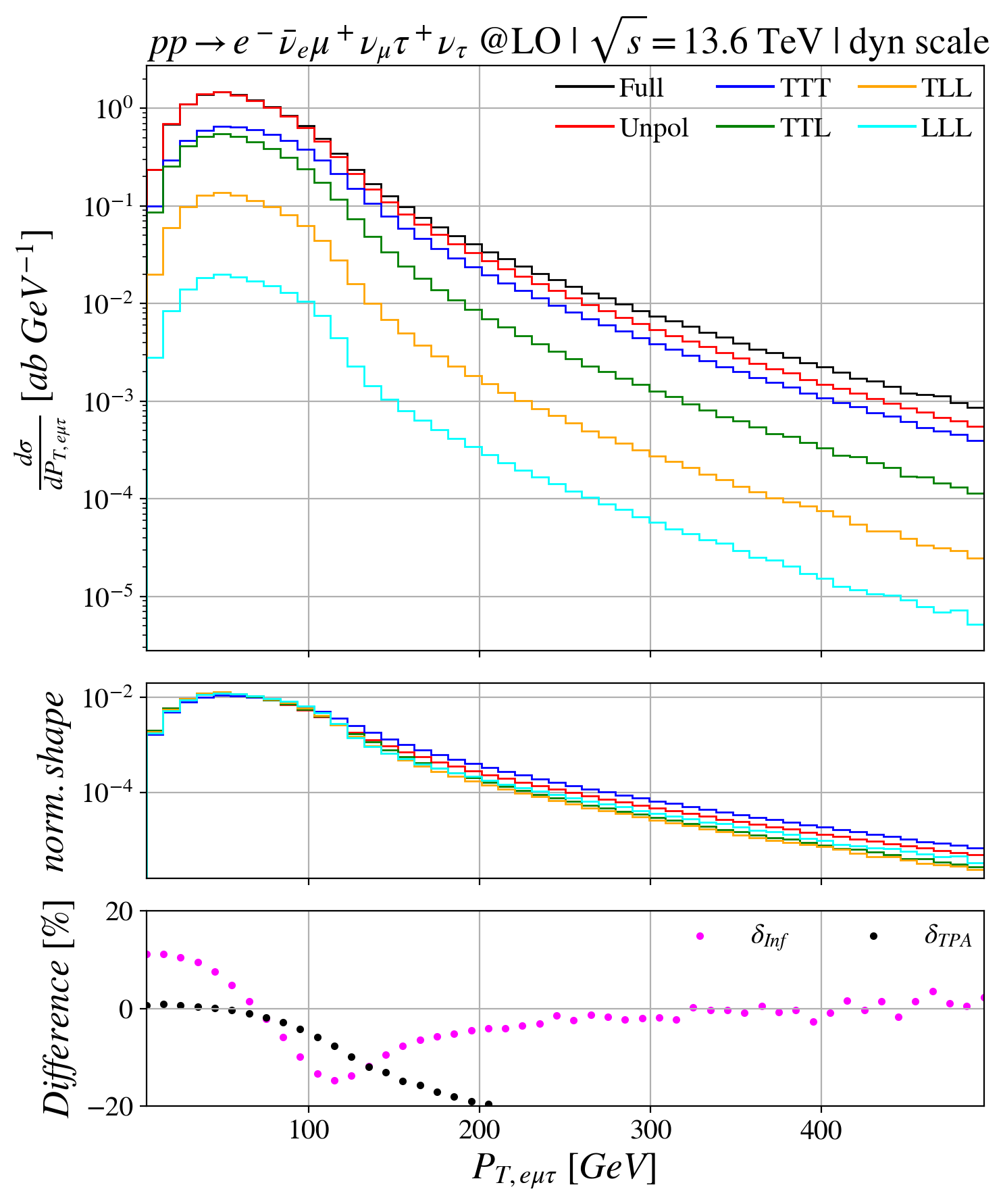} 
	\caption{Same as \fig{dist:thetaE_DeltaPhi_MT} but for the invariant mass of the six-lepton system (left) and the transverse momentum of the electron-muon-tauon system (right).} 
	\label{dist:m6l_pT_emt}
\end{figure}
We now present results for some kinematic distributions, which are important for the measurement of polarized cross sections. 
They are displayed in \fig{dist:thetaE_DeltaPhi_MT}, \fig{dist:y_theta_emt}, \fig{dist:m6l_pT_emt} for the $W^-W^+W^+$ process. 
Corresponding results for the $W^+W^-W^-$ process are shown in \fig{dist:thetaE_DeltaPhi_MT_pmm}, \fig{dist:y_theta_emt_pmm}, \fig{dist:m6l_pT_emt_pmm}. 
All kinematic observables ($x$ axis) are calculated in the Lab frame, except for the $\cos \theta_e^{WWW}$ 
in \fig{dist:thetaE_DeltaPhi_MT}. This polar angle is calculated as the angle between $\vec{p}^e_{{W^-}\text{-rest}}$ (momentum direction 
of the electron measured in the parent $W^-$ rest frame) and $\vec{p}^{W^-}_{WWW\text{-cmf}}$ (momentum of the $W^-$ in 
the $WWW$ c.m.f.). It is important to calculate $\vec{p}^e_{{W^-}\text{-rest}}$ from $\vec{p}^e_{\text{Lab}}$ via two boosts: first boost 
from the Lab frame to the $WWW$ c.m.f., second boost from the $WWW$ c.m.f. to the $W^-$ rest frame, as we are calculating the electron angle with respect to 
the $W$ direction in the $WWW$ c.m.f.. Boosting $p_e$ directly from the Lab frame to 
the $W^-$ rest frame would give a different direction. 

For template fits of polarized cross sections, one has to rely on a distinguished shape of a given polarization. 
The $\cos \theta_e^{WWW}$ distribution in \fig{dist:thetaE_DeltaPhi_MT} (left) gives distinctly different shapes for the four polarizations, 
the $\Delta\phi(\mu,\tau)$ in \fig{dist:thetaE_DeltaPhi_MT} (right) can distinguish clearly the TTT.
The $|\Delta y(\mu, e\tau)|$ and $\cos\theta(\tau, e\mu)$ distributions in \fig{dist:y_theta_emt} help to classify the four polarizations into two shapes, with 
one category for the TLL and LLL the other for the TTT and TTL. These kinds of angular distributions are therefore very helpful for 
the measurement of polarized cross sections.

The energy dependence of the polarized cross sections are plotted in \fig{dist:m6l_pT_emt}. 
For the $M_{6l}$ distribution (left), one would naively expect that, as the energy grows, the slope of the curve would increase with 
the number of longitudinal modes. This is clearly so for the cases of TTT, TTL, and TLL, with significant changes. 
However, the change from TLL to LLL is very mild. It is also interesting to notice the peak position of the differential cross sections. 
As the $M_{6l}$ approaches the triple-resonance threshold of $3M_W \approx 241$ GeV, looking at the normalized-shape panel, one sees that the 
LLL rises fastest, followed by the TLL, TTL, and TTT. The peak position of the LLL is therefore a bit shifted to the left of the threshold, while 
it is shifted to the right of the threshold for the TTT. The interference between the four polarization combinations is highest around the threshold, being 
about $5\%$ (purple line in the bottom panel), and then drops as the energy increases. 

Finally, the transverse momentum distribution of the $e-\mu-\tau$ system is plotted in \fig{dist:m6l_pT_emt} (right). At LO, this is identical to 
the missing transverse momentum distribution. As the transverse momentum grows, all polarized cross sections 
reach the peak at around $50$ GeV and then go down slowly. Then, at around $120$ GeV, the 
LLL cross section drops sharply. This also happens to the TLL and TTL, but with a lesser extent. 
These sudden jumps are due to the $m_{W^+W^-}$ cuts.

This remarkable different behavior of the LLL differential cross section is an indication of the usefulness of the LLL 
polarization. It can help to better understand the underlying dynamics of a phenomenon or, as one can imagine, to fit a model 
parameter (e.g. $M_W$) when there is a strong sensitivity. Before deciding on its application, one should however include radiative corrections to 
see whether this distinct behavior is smeared out. 

The large differences between the unpolarized TPA and the full off-shell cross sections (see the black line in the bottom panel) 
at large $P_{T,{e\mu\tau}}$ have been observed in \cite{Dittmaier:2019twg}. As pointed out there, 
this comes from two kinematic configurations. In one case, the large $P_{T,\text{miss}}$ is balanced against a single charged-lepton $P_{T,\ell}$ 
with the $P_T$ of the remaining charged leptons being soft. 
In the other case, the large $P_{T,e\mu\tau}$ recoils against a single neutrino transverse momentum while the other neutrinos are soft. 
These mechanisms are illustrated by the Feynman diagram in \fig{Feyn_dia} (g, h), classified as a single-resonance contribution. 

Concerning the comparison between the fixed and dynamical scales, integrated results in \tab{tab:mpp_mu0} show small differences. 
This is also the case for flat distributions such as angular distributions, where the deviations are within $5\%$ across the whole range. 
For other distributions such as $P_{T,\ell}$, $M_{\ell \ell^\prime}$, $\Delta y(\mu, e\tau)$, etc the differences can reach the level 
of tens of percent in the regions where the differential cross section is suppressed. 
These differences should be taken into consideration when comparing theoretical predictions with data.
\section{Comparisons with other tools}
\label{sect:compa}
In this section we compare our results to those of MoCaNLO-1.0.0 \cite{Denner:2026phn} and Sherpa-3.0.1 \cite{Sherpa:2024mfk,Hoppe:2023uux}. 
These comparisons are done using the fixed factorization scale of $\mu = 3M_W/2$ for simplicity.

Before discussing details of the comparisons, we first note the differences in the definition of the polarization combinations between the 
three programs. These differences occur due to the output format and can easily be fixed by some small changes in the source codes. 
As we are not familiar with the MoCaNLO and Sherpa source codes, we refrain from doing it here and just take all outputs of these programs 
to build the polarized cross sections most close to our definitions. This is good enough because the mismatches are some small interferences. 
The precise definitions of the polarized cross sections of the three programs read as follows.
\begin{itemize}
\item LLL: same for all programs as there is no interference.
\item TLL: MulBos includes the sum of $W^-_T W^+_L W^+_L$, $W^-_L W^+_T W^+_L$, $W^-_L W^+_L W^+_T$ combinations
and their interferences while MoCaNLO's results do not include the interferences. 
For Sherpa, it is the incoherent sum of $W^-_P W^+_L W^+_L$, $W^-_M W^+_L W^+_L$, $W^-_L W^+_P W^+_L$, $W^-_L W^+_M W^+_L$, 
$W^-_L W^+_L W^+_P$, $W^-_L W^+_L W^+_M$ cross sections, where the subscripts $P$ and $M$ stand for right-transverse and left-transverse modes, respectively. 
Thus, Sherpa's results do not include any interferences, MoCaNLO includes 
the right-transverse and left-transverse interferences, and MulBos includes all interferences.
\item TTL: MulBos includes the coherent sum of $W^-_T W^+_T W^+_L$, $W^-_T W^+_L W^+_T$, $W^-_L W^+_T W^+_T$ combinations, while 
MoCaNLO is just the incoherent sum of those three combinations. Sherpa's cross section reads
\bea
\sigma_{TTL}^\text{Sherpa} = \sum_{i,j\in [P,M]} (\sigma_{W^-_i W^+_j W^+_L} + \sigma_{W^-_i W^+_L W^+_j} + \sigma_{W^-_L W^+_i W^+_j}),
\eea
which is just the incoherent sum of $12$ polarized cross sections.
\item TTT: Sherpa's cross section reads
\bea
\sigma_{TTT}^\text{Sherpa} = \sum_{i,j,k\in [P,M]} \sigma_{W^-_i W^+_j W^+_k},
\eea
which is just the incoherent sum of $8$ polarized cross sections, while MulBos or MoCaNLO results are the 
coherent sum.
\end{itemize}
In summary, for the LLL polarization definition all three codes agree; for the TTT MulBos and MoCaNLO are the same while Sherpa 
differs by the left-transverse and right-transverse interferences; for the TLL and TTL MulBos includes full interferences, 
Sherpa has no interferences, while MoCaNLO contains partial interferences.
\begin{table}[h!]
	\renewcommand{\arraystretch}{1.3}	
	\begin{bigcenter}	 
        \setlength\tabcolsep{0.08cm}
	{\fontsize{8.0}{8.0}
	\begin{tabular}{|c|c|c|c|c|c|c|}
		\hline
		Code &  Off-shell & Unpol. & TTT & TTL & TLL & LLL \\ 
		\hline
		{\fontsize{8.0}{8.0}MulBos (SSOP-TPA)}     & 122.29(2) & 118.38(2)  & 60.75(1)   & 43.28(1) & 10.581(2) & 1.5998(4) \\
		{\fontsize{8.0}{8.0}MoCaNLO (SOP-TPA)} & 122.32(3)  & 118.53(3) & 60.75(1)  & 43.32(1) & 10.613(2)     & 1.5781(3)\\	
		{\fontsize{8.0}{8.0}Sherpa (NWA)} & 122.36(2)   & 120.60(7)   & 60.12(6)  & 43.29(5) & 10.63(2)     & 1.603(7)    \\				
		\hline
	\end{tabular}
	}
	\caption{Comparisons between MulBos, MoCaNLO, and Sherpa for the unpolarized and triply-polarized cross sections of the 
	$W^- W^{+} W^{+}$ process with the fixed factorization scale of $\mu = 3M_W/2$. 
	The cross section values are given in attobarn (ab). Note: polarized cross sections of MulBos, MoCaNLO, and Sherpa differ by some interferences, see text for more details.}
	\label{compa_tools}
	\end{bigcenter}	 
\end{table}
\begin{figure}[h!]
	\centering
		\includegraphics[width=0.49\textwidth]{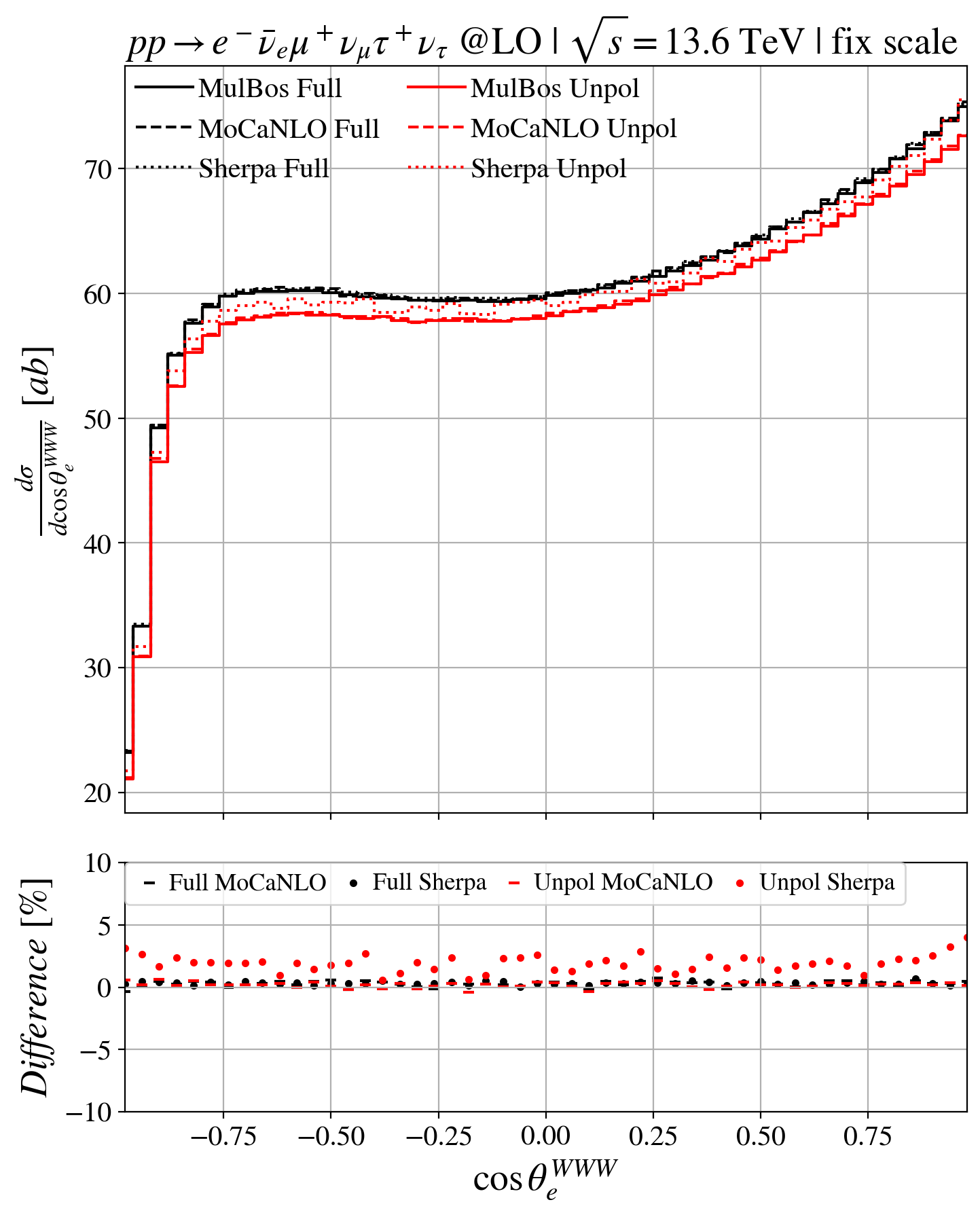}
		\includegraphics[width=0.49\textwidth]{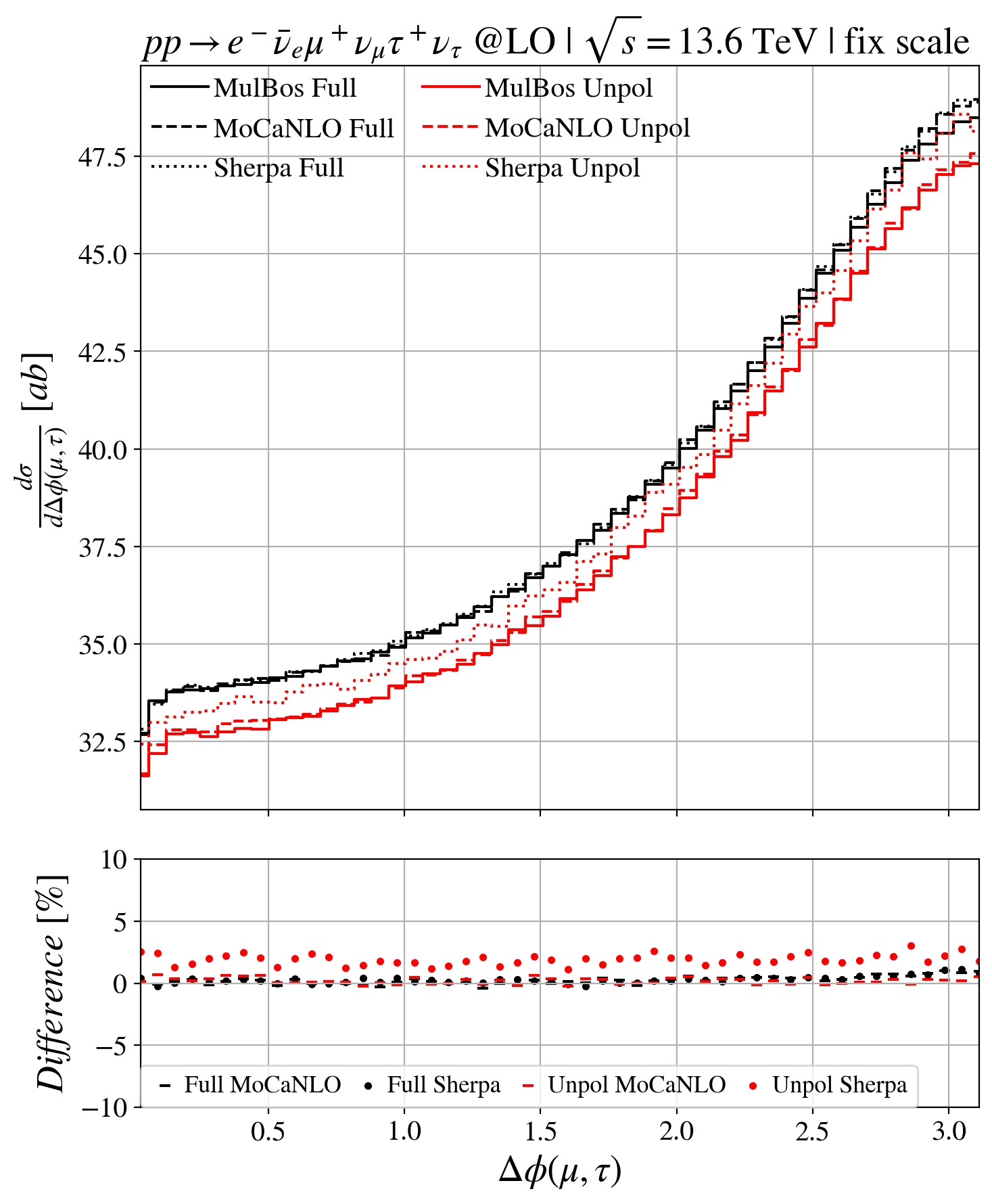} 
	\caption{Comparisons between Sherpa, MoCaNLO, and MulBos for the distributions in $\cos \theta_e^{WWW}$ (left) and 
	in $\Delta\phi(\mu,\tau)$ (right) with the fixed factorization scale of $\mu = 3M_W/2$. 
	Results for the full off-shell case and unpolarized pole approximations are shown. In the small panels, relative differences 
	compared to MulBos are plotted. Note: polarized cross sections of MulBos, MoCaNLO, and Sherpa differ by some interferences, see text for more details.} 
	\label{dist:thetaE_DeltaPhi_MT_compa_1}
\end{figure}
\begin{figure}[h!]
	\centering
		\includegraphics[width=0.49\textwidth]{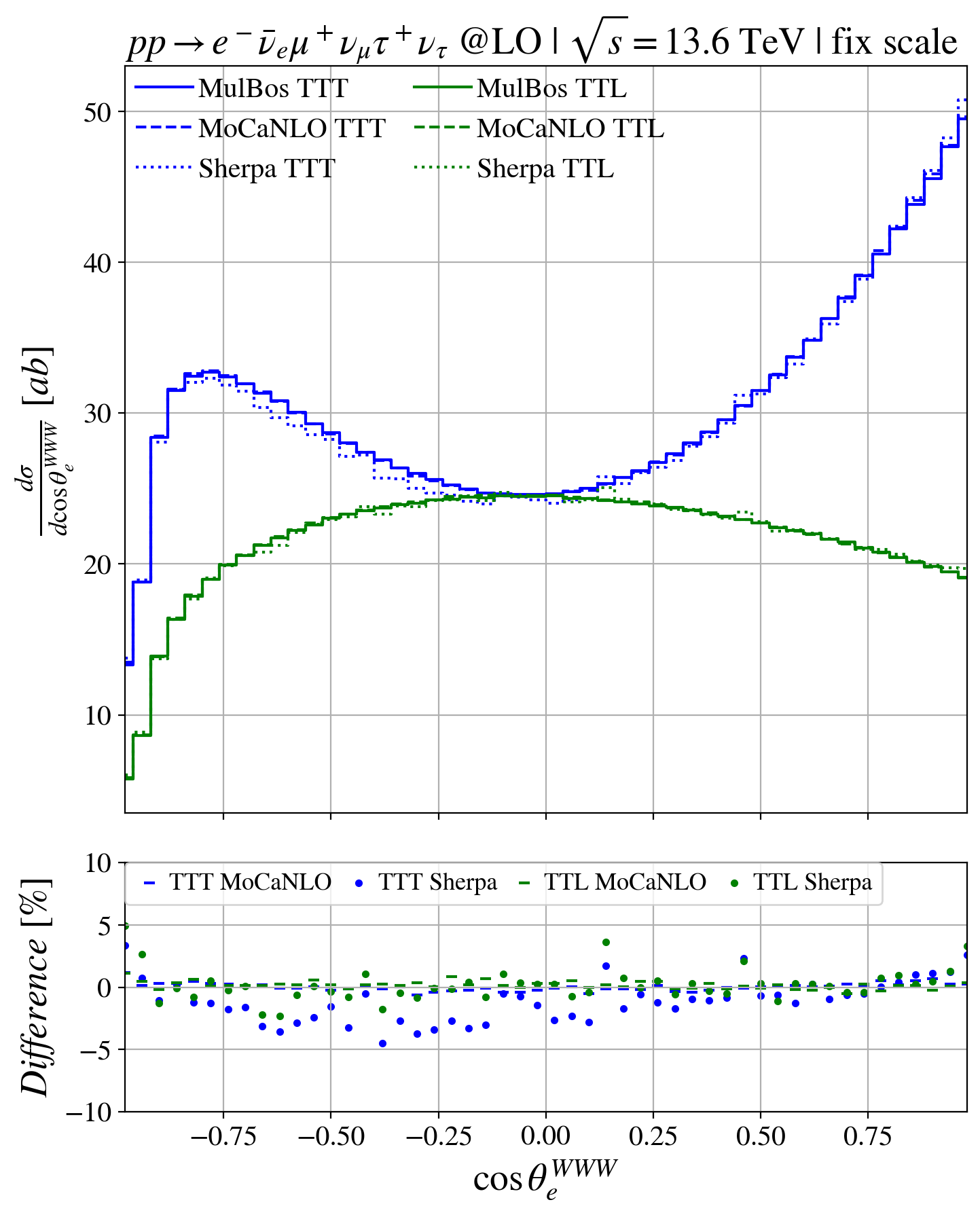}
		\includegraphics[width=0.49\textwidth]{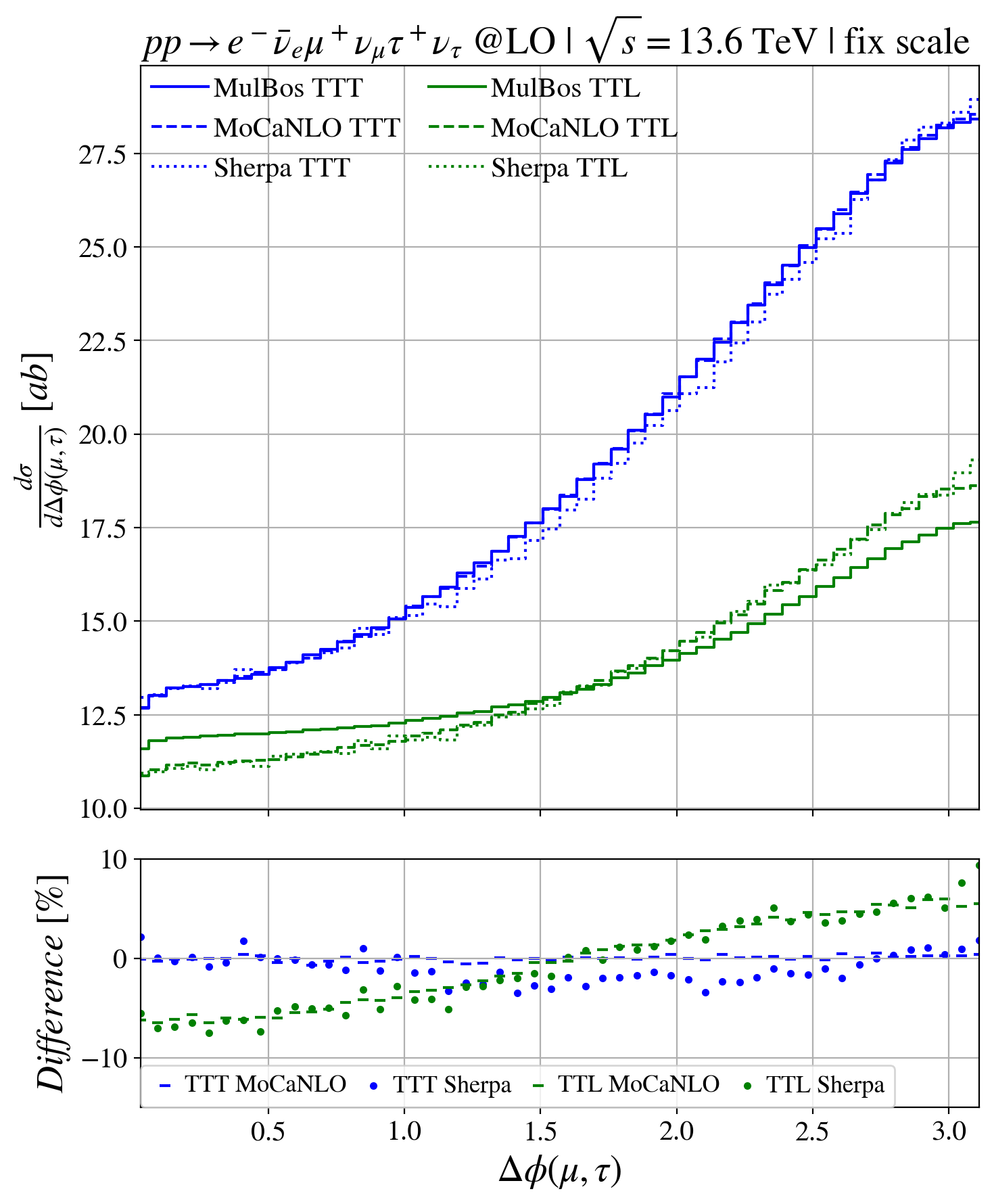} 
	\caption{Same as \fig{dist:thetaE_DeltaPhi_MT_compa_1} but for the TTT and TTL polarized cross sections.} 
	\label{dist:thetaE_DeltaPhi_MT_compa_2}
\end{figure}
\begin{figure}[h!]
	\centering
		\includegraphics[width=0.49\textwidth]{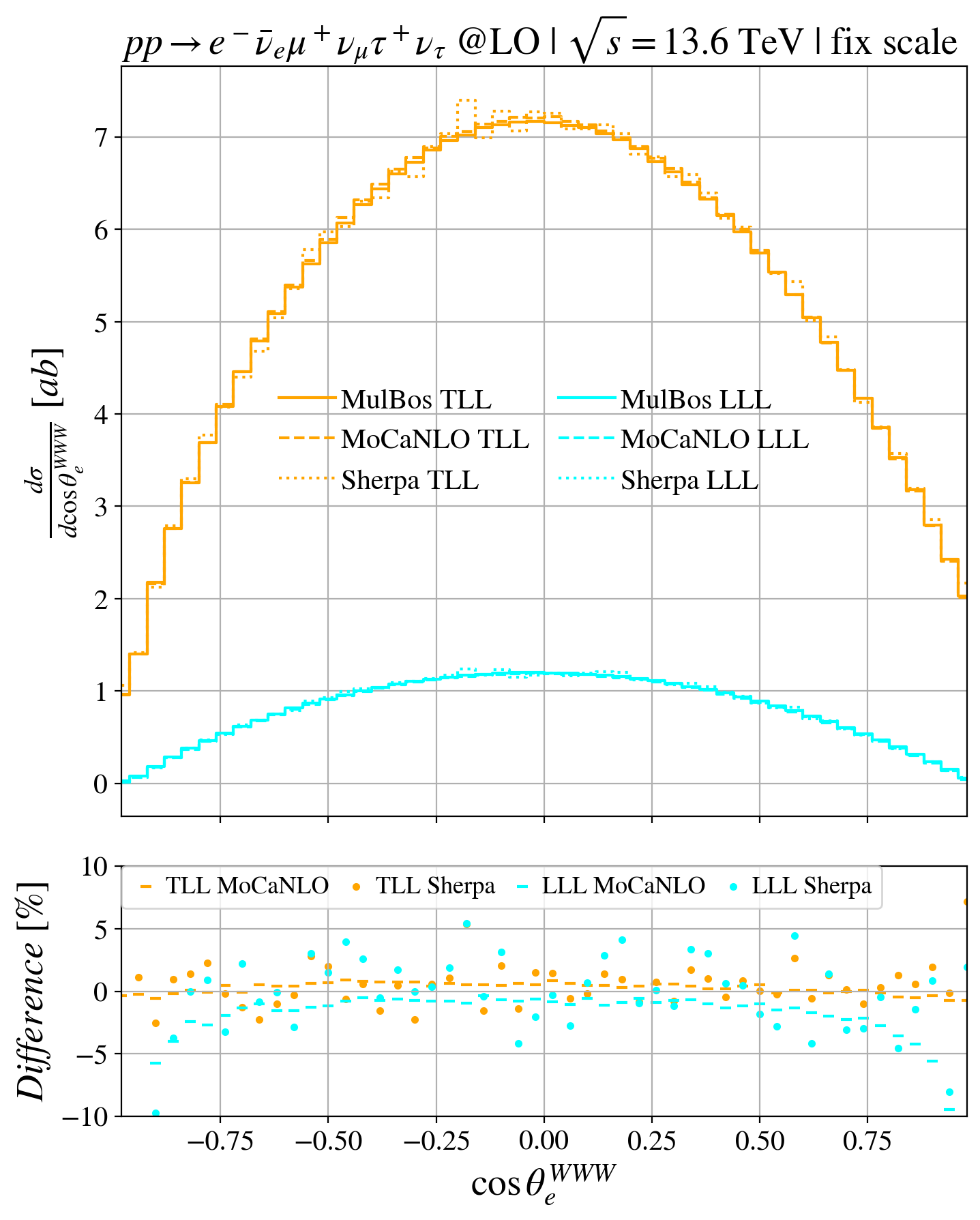}
		\includegraphics[width=0.49\textwidth]{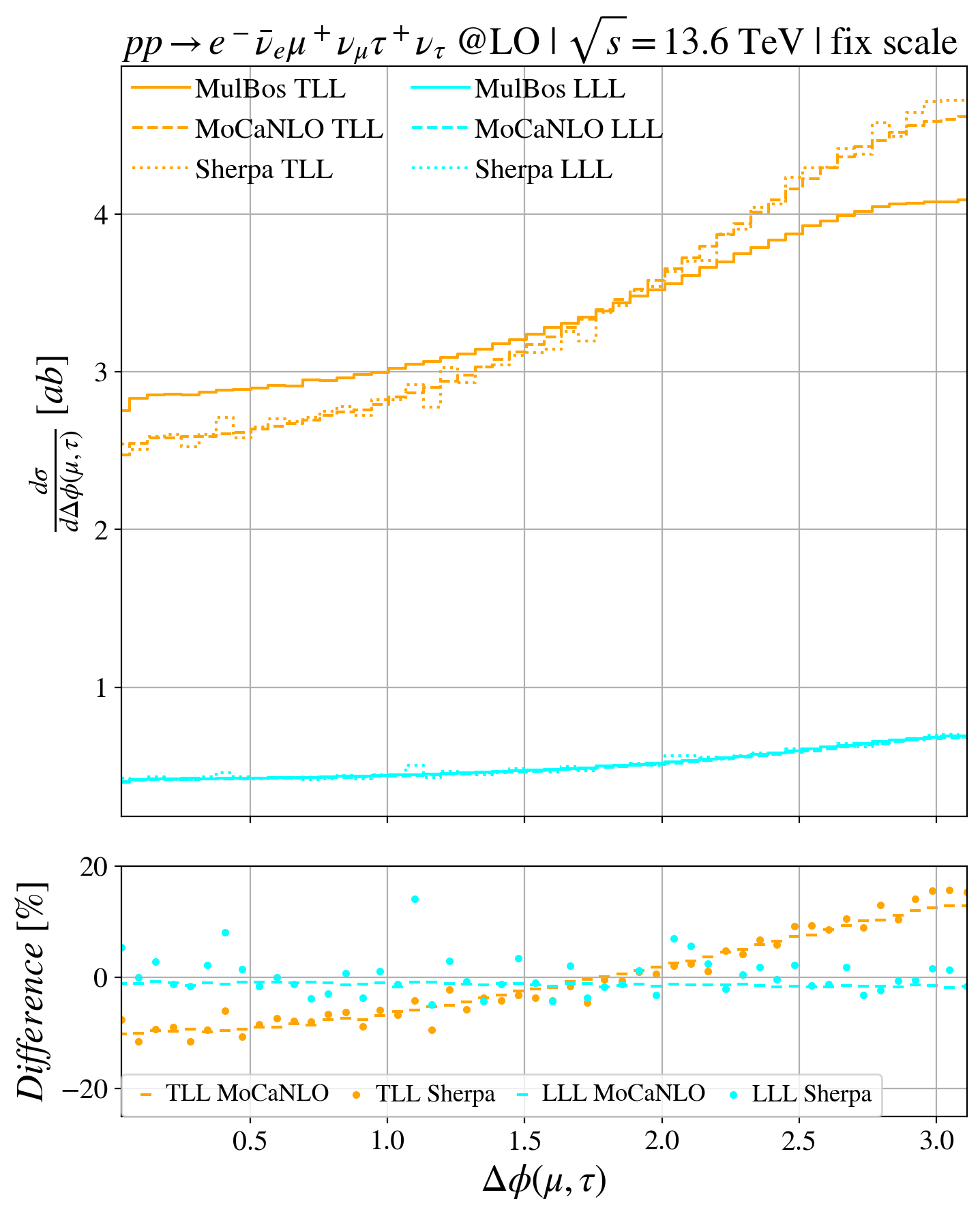} 
	\caption{Same as \fig{dist:thetaE_DeltaPhi_MT_compa_1} but for the TLL and LLL polarized cross sections.} 
	\label{dist:thetaE_DeltaPhi_MT_compa_3}
\end{figure}

We next discuss the differences in the calculation of the polarized cross sections. While MulBos and MoCaNLO use the 
TPA, Sherpa uses the NWA. The only difference between MulBos and MoCaNLO is at the on-shell projection. 
MulBos uses the SSOP while MoCaNLO the SOP. 
For the Sherpa run, the Breit-Wigner smearing is on to better model the shapes of the three resonances. The total widths of 
the $W$ bosons for this smearing match the value given in \sect{sec:input}. The $W$ partial decay widths to leptons are calculated 
at LO. 

With those differences in mind, we are now ready to discuss the integrated cross sections presented in \tab{compa_tools}. 
For the full off-shell case, all three programs agree very well with differences less than one per mil. 
The difference between MoCaNLO and MulBos is about $+0.1\%$ for the unpolarized TPA, approximately $0.0\%$ for the TTT, 
$+0.1\%$ for the TTL, $+0.3\%$ for the TLL, and $-1.4\%$ for the LLL, which are reasonable given the different on-shell mappings. 
The differences between Sherpa and MulBos read $+1.9\%$ for the unpolarized PA, 
$-1.0\%$ for the TTT, $+0.02\%$ for the TTL, $+0.5\%$ for the TLL, and $+0.2\%$ for the LLL. 
These differences are even smaller than the differences between Sherpa and MulBos for the $ZZ$ production reported in 
\cite{Carrivale:2025mjy}. Given that Sherpa uses the NWA while MulBos uses TPA, those small differences are acceptable.

We now discuss the kinematic distributions shown in \fig{dist:thetaE_DeltaPhi_MT_compa_1}, \fig{dist:thetaE_DeltaPhi_MT_compa_2}, 
and \fig{dist:thetaE_DeltaPhi_MT_compa_3}. The $\cos\theta_e^{WWW}$ distributions are in the left plots, 
while the $\Delta\phi(\mu,\tau)$ ones are in the right plots. In the big panels, the solid lines are MulBos, dashed lines are MoCaNLO, 
while dotted lines are Sherpa. In the small panels, the relative differences with respect to MulBos are plotted. 

For the full off-shell and unpolarized PA differential cross sections in \fig{dist:thetaE_DeltaPhi_MT_compa_1}, all differences 
are within $5\%$ cross the whole angular ranges. MoCaNLO results agree better with MulBos results than Sherpa ones. 

For the TTT and TTL combinations in \fig{dist:thetaE_DeltaPhi_MT_compa_2}, the agreement between MoCaNLO and MulBos is still excellent 
for the $\cos\theta_e^{WWW}$ distributions, while the differences between Sherpa and MulBos increase slightly in the region 
around $\cos\theta_e^{WWW} \approx -0.4$. The $\Delta\phi(\mu,\tau)$ distributions are interesting since they are sensitive to interference effects. 
We see a clear deviation between MoCaNLO and MulBos for the TTL case, while excellent agreement for the TTT combination. 
This is mainly because MoCaNLO and MulBos differ by some interferences for the TTL polarization. The effect of different on-shell 
mappings is very small, in our opinion. Similarly, the differences between Sherpa and MulBos are due to the interference mismatches 
and the difference between the NWA and the TPA. 

Finally, the TLL and LLL cases are plotted in \fig{dist:thetaE_DeltaPhi_MT_compa_3}. 
The comparison between MoCaNLO and MulBos for the $\cos\theta_e^{WWW}$ distributions are interesting. 
The agreement for the TLL is very good as this distribution is insensitive to interferences. 
For the LLL, the difference has a clear shape and increases up to $-10\%$ at the edges. As the LLL 
definition is the same in both programs, this difference is due to the different on-shell mappings. 
Similarly, the difference between Sherpa and MulBos for the LLL distribution is due to the 
difference between the NWA and the TPA. For the TLL case, it seems that the additional effect from the interferences 
cancels out the deviation between the NWA and the TPA, resulting in small differences between Sherpa and MulBos.

For the $\Delta\phi(\mu,\tau)$ distributions in the right plots, we see that the agreement between MoCaNLO and MulBos 
is very good for the LLL case, signaling that this distribution is not sensitive to the on-shell mapping. 
The large differences for the TLL, ranging from $-10\%$ to more than $+10\%$, are mainly due to the mismatched interferences. 
The differences between Sherpa and MulBos are ranging from $-10\%$ to $+20\%$ for the TLL, while being mostly within $10\%$ for the LLL.
\section{Conclusions}
\label{sec:con}
We have presented, for the first time, all triply-polarized cross sections for the $WWW$ production processes at the LHC. 
Results were obtained at LO with fully leptonic decays. 

We found that the LLL cross section is very small at the integrated or differential cross section level. 
It is therefore very challenging to measure it unless higher order corrections or/and new physics
increase it to the level of tens of percent. However, the suppressed value of the SM LLL cross section 
should not discourage us from performing its measurement as an upper limit would be valuable 
for constraining new-physics parameters such as the anomalous gauge couplings.

It was also found that the polarization interference is of the same size as the LLL contribution. A separate template for the interference 
would then be needed in the measurement. 

Compared to the diboson processes, a new ingredient for triboson processes is the on-shell mapping. 
We have presented a new and general method for $n$-resonance processes, named the sequential splitting on-shell 
projection. An essential idea is a dynamic and democratic mapping. 
Compared to the sequential pairing method proposed by Dittmaier and Schwan where 
the TPA conditions are hidden, our mapping shows clearly the TPA conditions. Numerically, the two methods give very similar results, though. 
Comparisons with the simultaneous projection of Denner, Lombardi, and Schwan also show excellent agreements. 
Moreover, good agreements with Sherpa, which uses the NWA, have been observed.

As a compensation for being small in cross section, triboson processes have numerous kinematic distributions with high 
power in shape distinction for different polarizations, in particular distributions in various angles between the flight 
directions of the charged leptons. Some representatives of those have been shown in this study. 
These distributions will be important for polarization measurements.  

The next natural step is to calculate NLO corrections to see how the LLL fraction and the interference change. This is still work 
in progress.
\newpage
\appendix
\section{Results for the $W^+W^-W^-$ process}
\label{appen_pmm}
\begin{figure}[h!]
	\centering
	\includegraphics[width=0.49\textwidth]{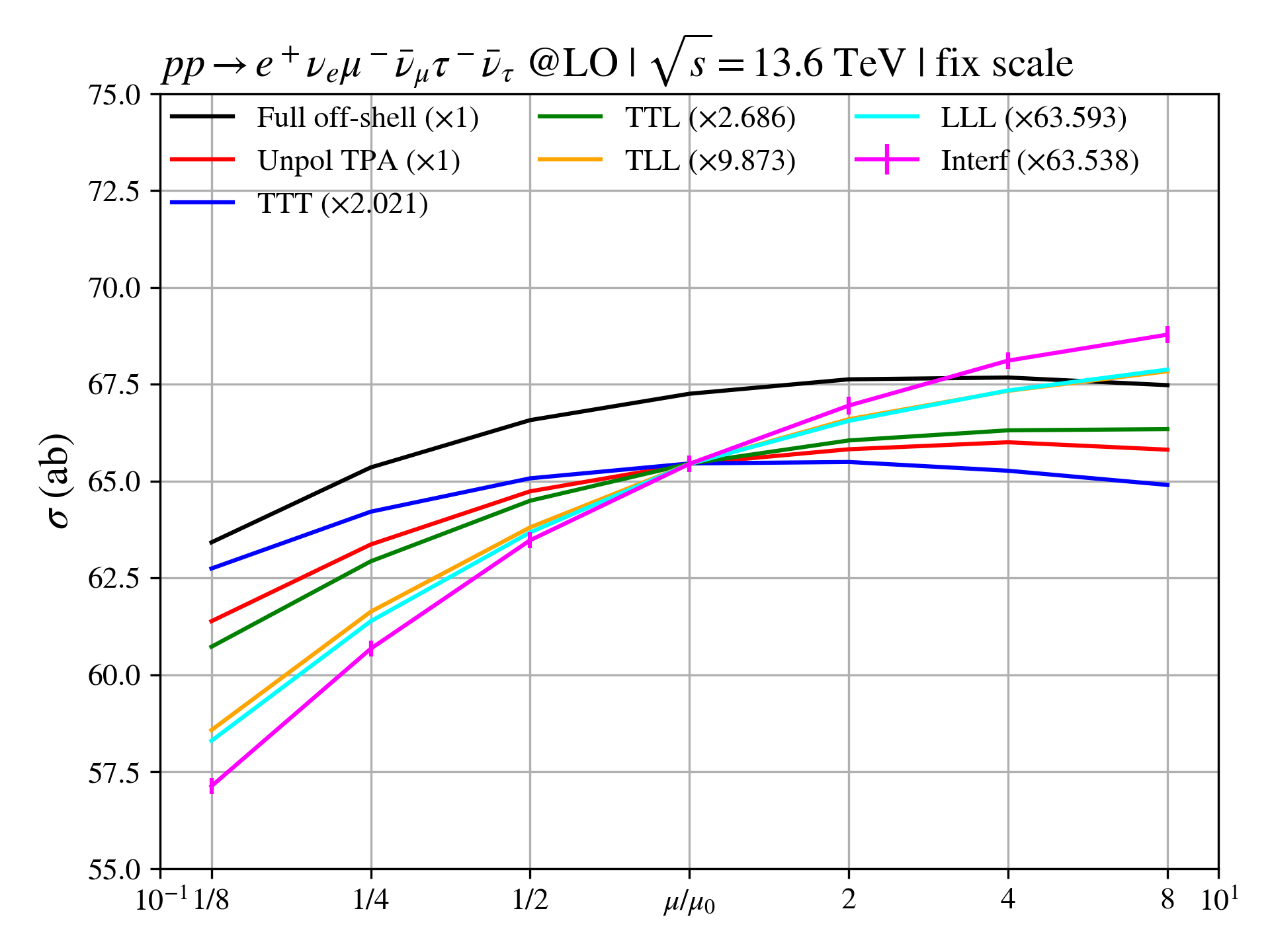}
	\includegraphics[width=0.49\textwidth]{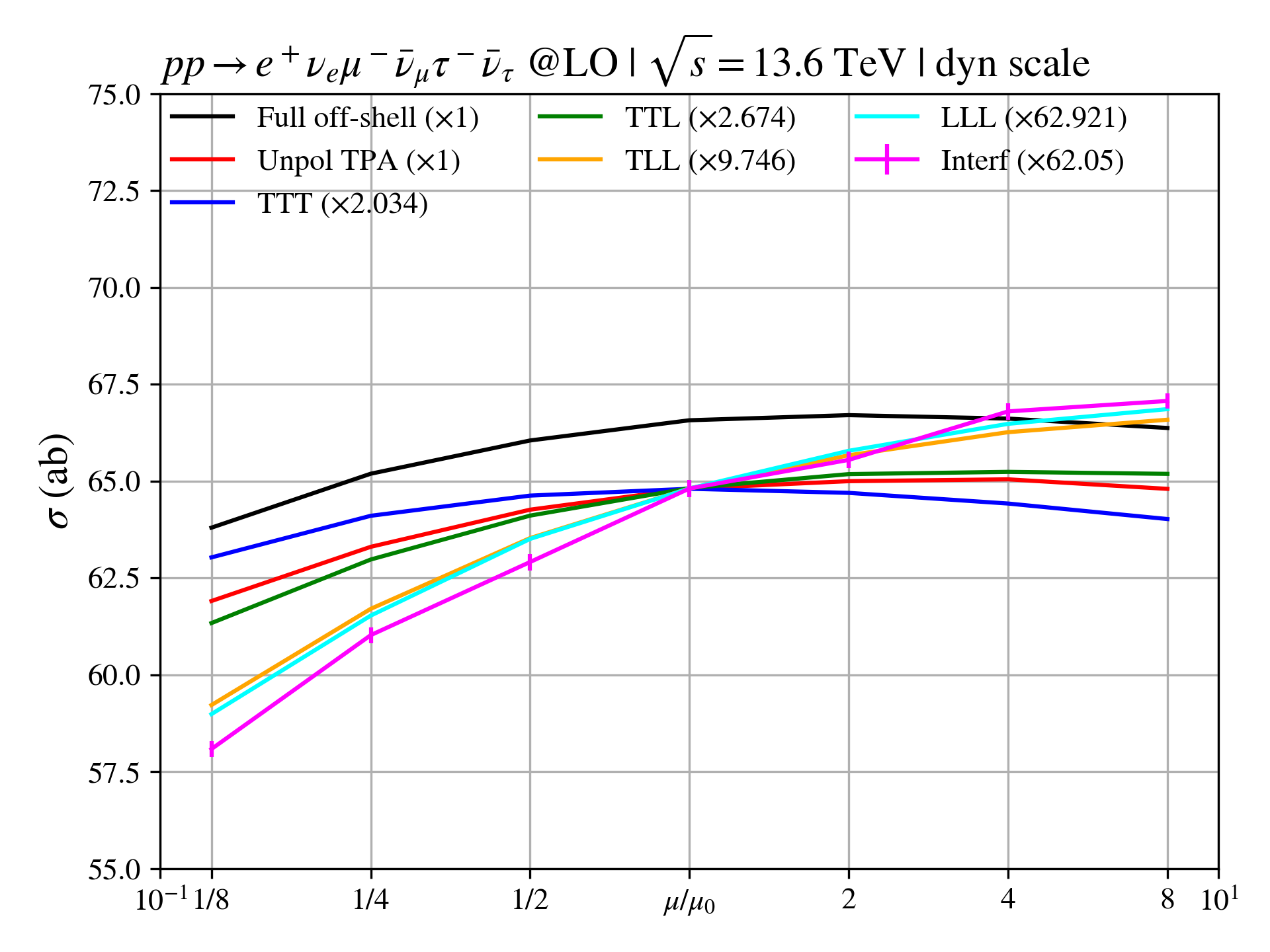} 
	\caption{Same as \fig{fig:xi_dep} but for the process $pp \rightarrow e^+ \nu_e \mu^- \bar{\nu}_\mu \tau^- \bar{\nu}_\tau + X$.} 
	\label{fig:xi_dep_pmm}
\end{figure}

\begin{figure}[h!]
	\centering
		\includegraphics[width=0.49\textwidth]{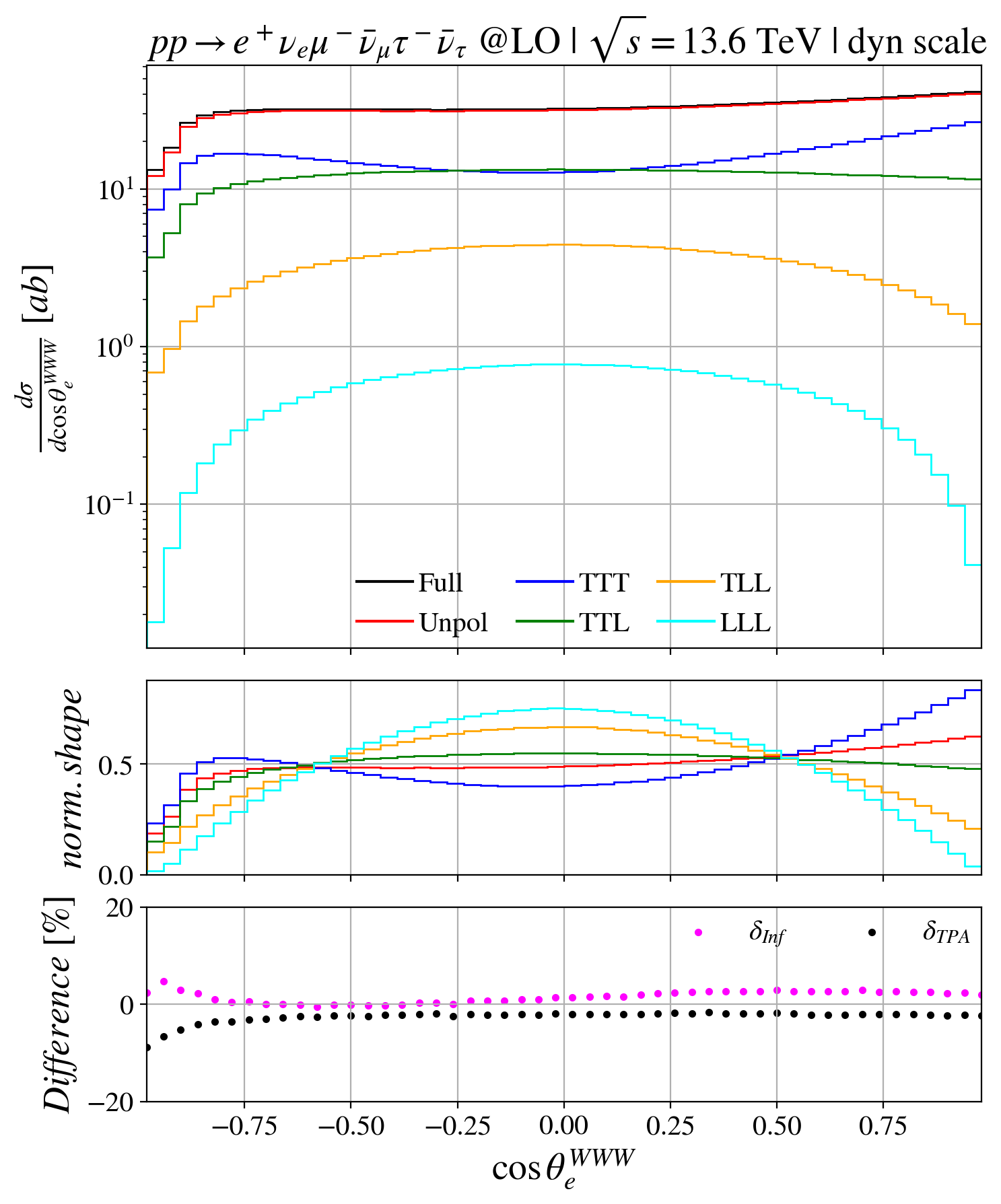}
		\includegraphics[width=0.49\textwidth]{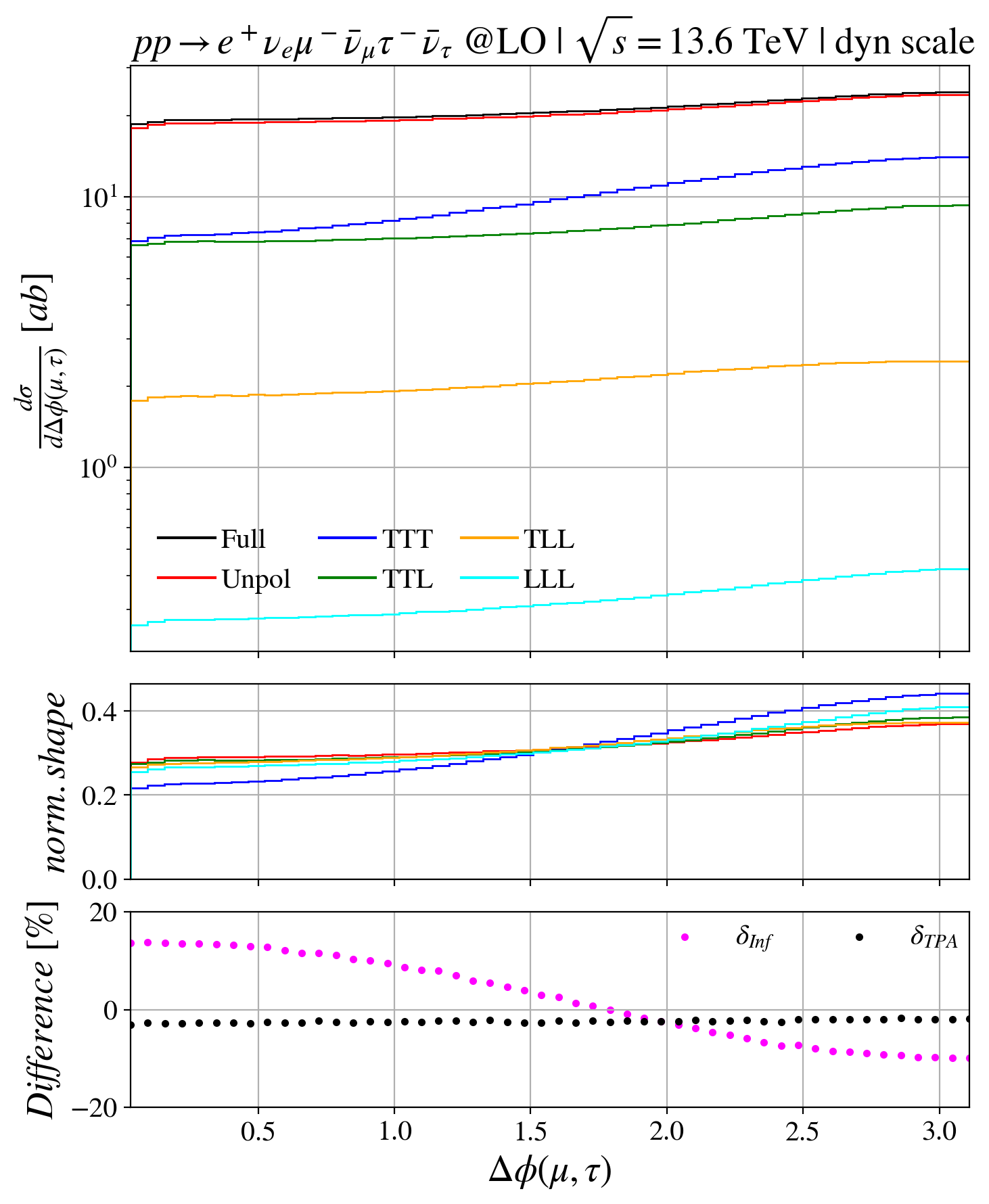} 
	\caption{Same as \fig{dist:thetaE_DeltaPhi_MT} but for the process $pp \rightarrow e^+ \nu_e \mu^- \bar{\nu}_\mu \tau^- \bar{\nu}_\tau + X$.}  
	\label{dist:thetaE_DeltaPhi_MT_pmm}
\end{figure}

\begin{figure}[h!]
	\centering
		\includegraphics[width=0.49\textwidth]{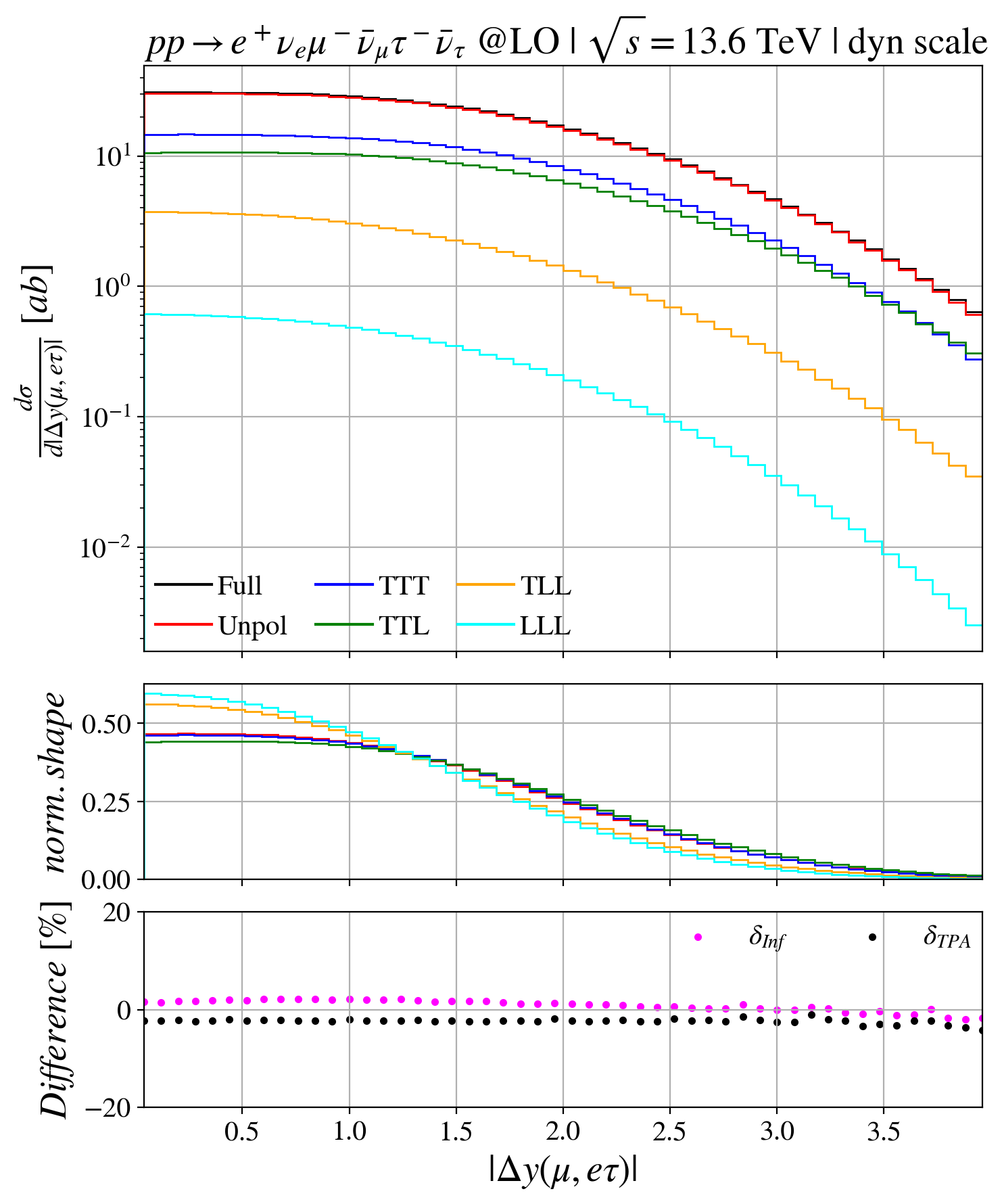} 
		\includegraphics[width=0.49\textwidth]{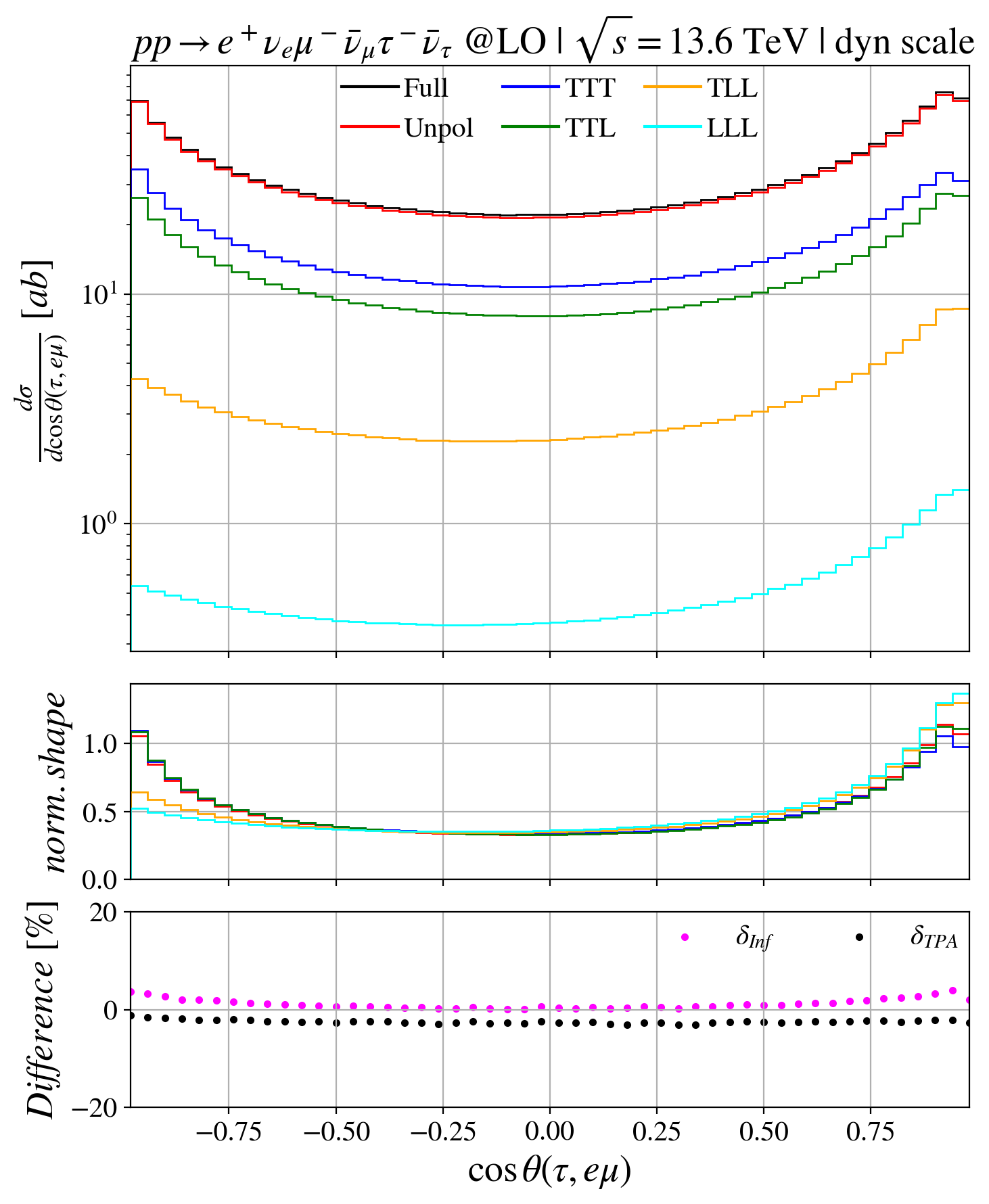}
	\caption{Same as \fig{dist:thetaE_DeltaPhi_MT} but for the rapidity difference between the muon and the electron-tauon system (left) and cosine of the angle 
	between the tauon and the electron-muon system (right) of the process $pp \rightarrow e^+ \nu_e \mu^- \bar{\nu}_\mu \tau^- \bar{\nu}_\tau + X$.} 
	\label{dist:y_theta_emt_pmm}
\end{figure}

\begin{figure}[h!]
	\centering
		\includegraphics[width=0.49\textwidth]{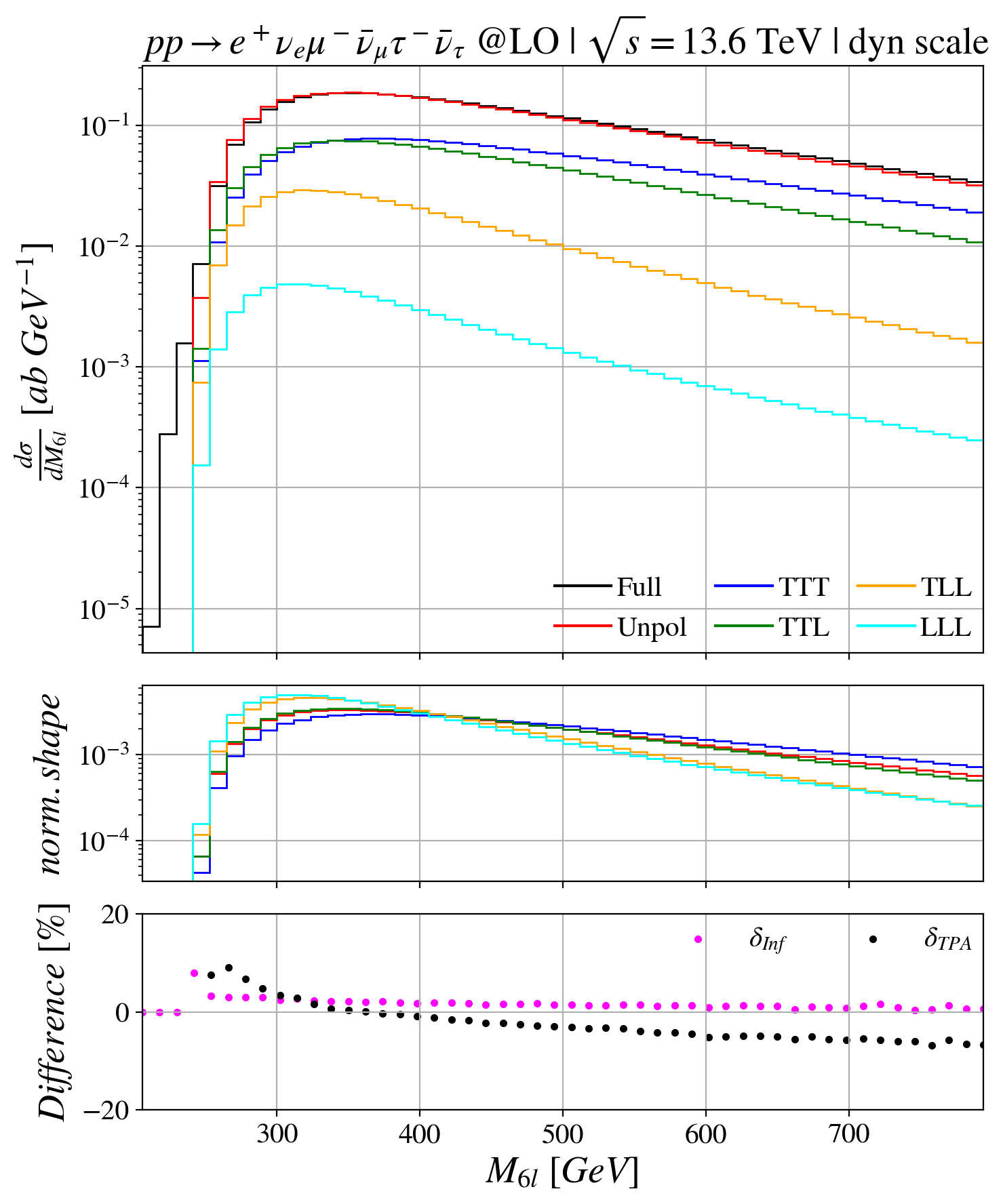}
		\includegraphics[width=0.49\textwidth]{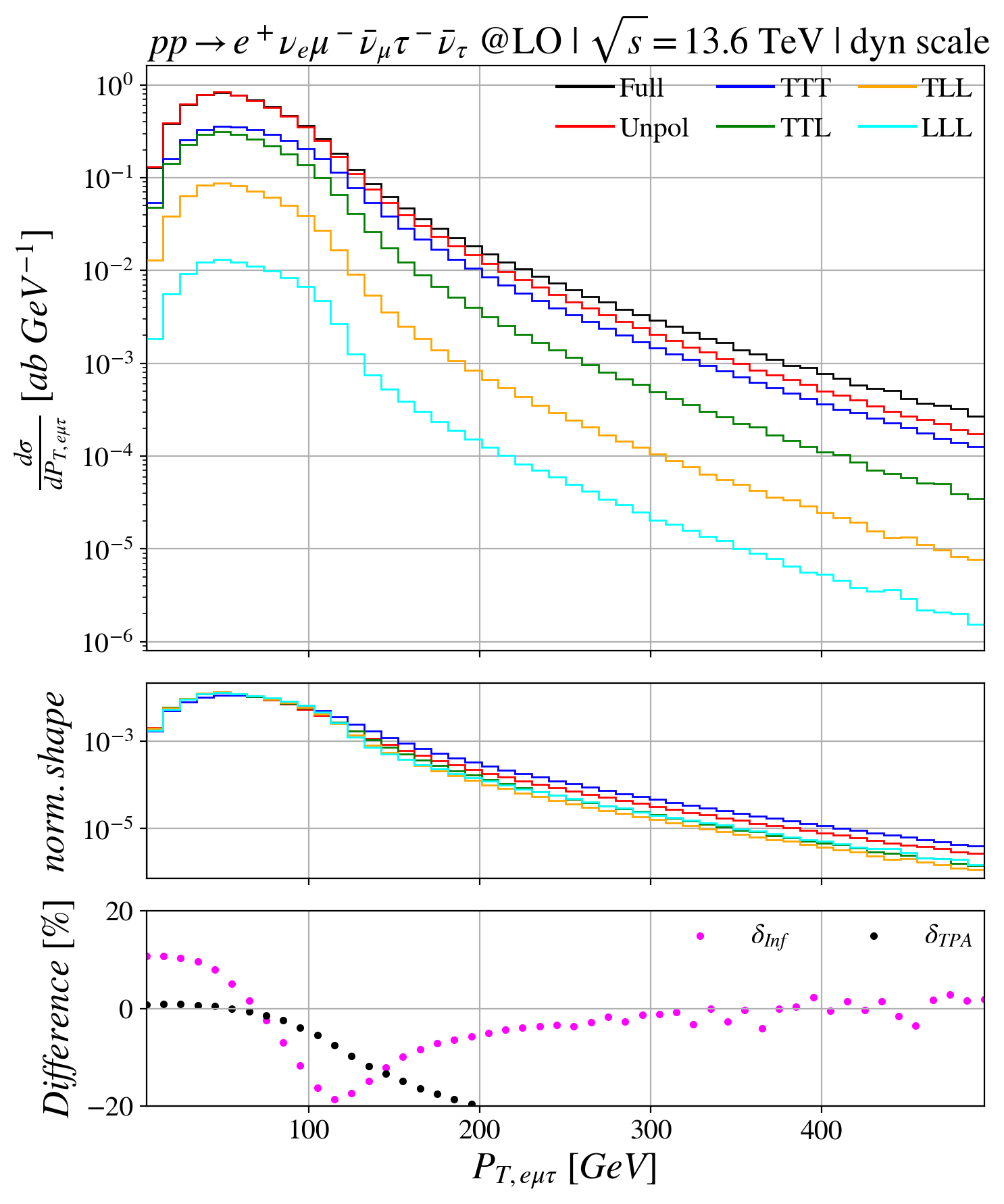} 
	\caption{Same as \fig{dist:thetaE_DeltaPhi_MT} but for the invariant mass of the six-lepton system (left) and the transverse momentum of the electron-muon-tauon system (right) 
	of the process $pp \rightarrow e^+ \nu_e \mu^- \bar{\nu}_\mu \tau^- \bar{\nu}_\tau + X$.} 
	\label{dist:m6l_pT_emt_pmm}
\end{figure}

In this appendix results for the process $pp \to e^+\nu_e \mu^-\bar{\nu}_{\mu} \tau^- \bar{\nu}_{\tau} + X$ (named $W^+W^-W^-$ for short) are provided. 
Integrated cross sections are given in \tab{tab:pmm_mu0} and \fig{fig:xi_dep_pmm}, while differential distributions are displayed in \fig{dist:thetaE_DeltaPhi_MT_pmm}, 
\fig{dist:y_theta_emt_pmm}, and \fig{dist:m6l_pT_emt_pmm}.
\begin{table}[ht!]
	\centering 
	\renewcommand{\arraystretch}{1.3}	
	\begin{bigcenter}	 
		{\fontsize{9.0}{9.0}
\begin{tabular}{|c|c|c|c|c|c|} \hline
	& {\fontsize{9.0}{9.0} $\sigma_{\text{fix}} ~[\text{ab}]$} & $f_{\text{fix}} [\%]$  & $\sigma_{\text{dyn}} [\text{ab}]$  & $f_{\text{dyn}} [\%]$ & $\delta_\sigma [\%]$ \\
        \hline
{\fontsize{9.0}{9.0}$\text{Full off-shell}$} & $67.293(8)^{+0.50\%}_{-1.06\%}$ & -- & $66.574(8)^{+0.20\%}_{-0.78\%}$ & -- & $+1.1$ \\
{\fontsize{9.0}{9.0}$\text{Unpol. TPA}$} & $65.51(1)^{+0.48\%}_{-1.17\%}$ & $100$ & $64.88(1)^{+0.19\%}_{-0.94\%}$ & $100$ & $+1.0$ \\
\hline
{\fontsize{9.0}{9.0}$\text{TTT}$} & $32.416(6)^{+0.00\%}_{-0.67\%}$ & $49.5^{+0.51\%}_{-0.51\%}$ & $31.887(6)^{+0.00\%}_{-0.35\%}$ & $49.1^{+0.60\%}_{-0.43\%}$ & $+1.7$ \\
{\fontsize{9.0}{9.0}$\text{TTL}$} & $24.384(5)^{+0.85\%}_{-1.53\%}$ & $37.2^{+0.36\%}_{-0.36\%}$ & $24.263(5)^{+0.47\%}_{-1.18\%}$ & $37.4^{+0.28\%}_{-0.24\%}$ & $+0.5$ \\
{\fontsize{9.0}{9.0}$\text{TLL}$} & $6.638(1)^{+1.63\%}_{-2.63\%}$ & $10.1^{+1.14\%}_{-1.47\%}$ & $6.657(1)^{+1.23\%}_{-2.07\%}$ & $10.3^{+1.03\%}_{-1.14\%}$ & $-0.3$ \\
{\fontsize{9.0}{9.0}$\text{LLL}$} & $1.0284(2)^{+1.77\%}_{-2.63\%}$ & $1.6^{+1.28\%}_{-1.47\%}$ & $1.0309(2)^{+1.42\%}_{-2.09\%}$ & $1.6^{+1.22\%}_{-1.16\%}$ & $-0.2$ \\
{\fontsize{9.0}{9.0}$\text{Interference}$} & $1.030(3)^{+2.30\%}_{-3.01\%}$ & $1.6^{+1.81\%}_{-1.86\%}$ & $1.045(4)^{+1.14\%}_{-2.93\%}$ & $1.6^{+0.95\%}_{-2.00\%}$ & $-1.4$ \\
\hline
\end{tabular}
		}
		\caption{Same as \tab{tab:mpp_mu0} but for the process $pp \rightarrow e^+ \nu_e \mu^- \bar{\nu}_\mu \tau^- \bar{\nu}_\tau + X$.}
		\label{tab:pmm_mu0}
	\end{bigcenter}	
\end{table}
\newpage
\acknowledgments
This research is supported by the NAFOSTED-DFG grant with the grant number DFG.103-2024.01. 
DNL thanks Ninh Van Thu for fruitful discussion on the on-shell projection. 
We are grateful to Mareen Hoppe for her help with the Sherpa comparison. 
Last but not least, we would like to thank the anonymous referee for a very helpful and detailed 
report which inspired us to improve significantly the paper. 
\newpage

\providecommand{\href}[2]{#2}\begingroup\raggedright\endgroup
\end{document}